\documentclass[10pt,twocolumn,superscriptaddress,english,10pt,prl,showpacs,floatfix,aps]{revtex4-1}
\usepackage[latin9]{inputenc}
\usepackage{amsthm}
\usepackage{amsmath}
\usepackage{amssymb}
\usepackage{esint}
\usepackage{graphicx}
\usepackage{upgreek}
\usepackage{xspace}
\usepackage{ulem}
\usepackage{CJK}

\makeatletter
\@ifundefined{textcolor}{}
{%
 \definecolor{BLACK}{gray}{0}
 \definecolor{WHITE}{gray}{1}
 \definecolor{RED}{rgb}{1,0,0}
 \definecolor{GREEN}{rgb}{0,1,0}
 \definecolor{BLUE}{rgb}{0,0,1}
 \definecolor{CYAN}{cmyk}{1,0,0,0}
 \definecolor{MAGENTA}{cmyk}{0,1,0,0}
 \definecolor{YELLOW}{cmyk}{0,0,1,0}
}

\usepackage{soul}
\usepackage{braket}
\usepackage[backref=false,
bookmarksnumbered=true,
bookmarks=true,
bookmarksopen=true,
colorlinks=true,
citecolor=blue,
linkcolor=blue,
anchorcolor=green,
urlcolor=blue,unicode=false]{hyperref}
\let\baraccent=\= % rename builtin command \= to \baraccent
\renewcommand{\=}[1]{\stackrel{#1}{=}} % for putting numbers above =

\makeatother

\usepackage{babel}

\begin{document}

\title{Magnetic anisotropy in surface-supported single-ion lanthanide complexes}


\author{Paul Stoll}
\affiliation{\mbox{Fachbereich Physik, Freie Universit\"at Berlin, 14195 Berlin, Germany}}

\author{Matthias Bernien}
\email{bernien@physik.fu-berlin.de}
\affiliation{\mbox{Fachbereich Physik, Freie Universit\"at Berlin, 14195 Berlin, Germany}}

\author{Daniela Rolf}
\affiliation{\mbox{Fachbereich Physik, Freie Universit\"at Berlin, 14195 Berlin, Germany}}

\author{Fabian Nickel}
\affiliation{\mbox{Fachbereich Physik, Freie Universit\"at Berlin, 14195 Berlin, Germany}}

\author{\begin{CJK}{UTF8}{bsmi}Qingyu Xu (\CJKfamily{gbsn}﻿徐庆宇)\end{CJK}}
\email{on leave of absence from Department of Physics, Southeast University, Nanjing 211189, P.\ R.\ China}
\affiliation{\mbox{Fachbereich Physik, Freie Universit\"at Berlin, 14195 Berlin, Germany}}

\author{Claudia Hartmann}
\affiliation{\mbox{Fachbereich Physik, Freie Universit\"at Berlin, 14195 Berlin, Germany}}

\author{Tobias R.\ Umbach}
\affiliation{\mbox{Fachbereich Physik, Freie Universit\"at Berlin, 14195 Berlin, Germany}}

\author{Jens Kopprasch}
\affiliation{\mbox{Fachbereich Physik, Freie Universit\"at Berlin, 14195 Berlin, Germany}}

\author{Janina N.\ Ladenthin}
\affiliation{\mbox{Fachbereich Physik, Freie Universit\"at Berlin, 14195 Berlin, Germany}}

\author{Enrico Schierle}
\affiliation{\mbox{Helmholtz-Zentrum Berlin f\"{u}r Materialien und Energie, 
12489 Berlin, Germany}}

\author{Eugen Weschke}
\affiliation{\mbox{Helmholtz-Zentrum Berlin f\"{u}r Materialien und Energie, 
12489 Berlin, Germany}}

\author{Constantin Czekelius}
\affiliation{\mbox{Institut f\"ur Organische Chemie und Makromolekulare Chemie, Heinrich-Heine Universit\"{a}t D\"{u}sseldorf, 40225 D\"usseldorf, Germany}}

\author{Wolfgang Kuch}
\affiliation{\mbox{Fachbereich Physik, Freie Universit\"at Berlin, 14195 Berlin, Germany}}

\author{Katharina J.\ Franke}
\affiliation{\mbox{Fachbereich Physik, Freie Universit\"at Berlin, 14195 Berlin, Germany}}

\date{\today}

\begin{abstract}
Single-ion lanthanide-organic complexes can provide stable magnetic moments with well-defined orientation  for spintronic applications on the atomic level. Here, we show by a combined experimental approach of scanning tunneling microscopy and X-ray absorption spectroscopy that dysprosium\--tris\-(1,1,1-trifluoro-4-(2-thienyl)-2,4\-butane\-dio\-nate) (Dy(tta)$_3$) complexes deposited on a Au(111) surface undergo a molecular distortion, resulting in distinct crystal field symmetry imposed on the Dy ion. This leads to an easy-axis magnetization direction in the ligand plane. Furthermore, we show that tunneling electrons hardly couple to the spin excitations, which we ascribe to the shielded nature of the $4f$ electrons.

\end{abstract}

\maketitle

\footnotetext{\P~on leave of absence from Department of Physics, Southeast University, Nanjing 211189, P.\ R.\ China}

\section{Introduction}
The challenge of modern spintronics is to store and process magnetic information and carry out quantum computation at the ultimately small limit \cite{Leuenberger2001,Bogani2008,Bertaina2008,Mannini2009,dei2011,Santini2011}. Single-molecule magnets (SMMs) may qualify as such units due to their slow relaxation of the magnetization~\cite{Sessoli1993,ishikawa03}. The most prominent cases of SMMs are multi-core $3d$ transition metal complexes with a large magnetic moment arising from the exchange-coupled paramagnetic ions~\cite{Lis80,caneschi91,Gatteschi00,Christou2000, beedle08, Milios2007}. Metal-organic complexes with only one $3d$ transition-metal ion typically do not qualify as SMMs, because they have a small magnetic moment and their anisotropy barrier is too small to allow for stabilizing a magnetic state for sufficiently long timescales. The reason for both these drawbacks can be ascribed to the participation of the $3d$ electrons in the metal-organic bond, which leads to quenching of the orbital momentum by mixing of $d$-states with opposite magnetic quantum number. Hence, the magnetic moment only arises from the spin and the spin-orbit coupling appears solely as a perturbation. This leads to small zero-field splittings and therefore to small anisotropy barriers for the magnetic moment.

Lanthanide single-ion complexes~\cite{ishikawa03,dreiserbrune, Thiele2014, Cucinotta12, Dreiser2016} are promising molecules to overcome the quenching of angular momentum. The $4f$ electrons of the lanthanides do not contribute to the bonding with the ligand since they exhibit only a small radial expansion and are shielded by the more delocalized $5s$ and $5p$ electrons~\cite{moeller1965}. Thus, the ligand field is only a perturbation and the total angular momentum remains a good quantum number~\cite{jensen1991rare}. The retained orbital momentum leads to a magnetic moment being larger than in $3d$ metal ions~\cite{abragambleaney}. Due to the large orbital moment, also the second ingredient, namely a large anisotropy barrier, may be found in such complexes, if a suitable crystal field is present.
Whereas the decoupling of the $4f$ electrons is important for protecting the magnetic state, it is a drawback for fixing the alignment of the anisotropy axis and for addressing and reading out the magnetic state \cite{Steinbrecher16}.
Hence, it is of interest how strongly $4f$ electrons can couple to electrons tunneling through individual atoms and molecules on surfaces \cite{Ternes09, Coffey15, Fahrendorf13, Komeda11, Warner16}.

Here, we show that dysprosium-tris(1,1,1-trifluoro-4- (2-thienyl) -2,4-butanedionate) (Dy(tta)$_3$) complexes are distorted from their anticipated gasphase structure by their adsorption on a metal surface. The resulting crystal field leads to an anisotropy that can be described by an easy axis of magnetization parallel to the surface, which is a direct consequence of the orientation of the Dy $4f$ orbitals. Whereas the shielding of the $4f$ electrons favors the large magnetic moment and anisotropy, our tunneling spectra indicate that it disfavors an electronic read-out of the magnetic state.

\section{Experimental}

\subsection{Deposition of the Complexes on Au(111)}

The Dy(tta)$_3$(H$_2$O)$_2$ complex was prepared from dysprosium chloride hydrate and the sodium dionate in aqueous solution and isolated as the dihydrate Dy(tta)$_3$(H$_2$O)$_2$ \cite{melby64,Purushotham1965182}. In a similar fashion, Gd(tta)$_3$(H$_2$O)$_2$ was prepared from gadolinium chloride hydrate. It is known that rare earth diketodionate complexes can lose water ligands under high-vacuum conditions~\cite{Binnemans2005107}. In addition, also partial hydrolysis of the diketodionate ligand may occur by formation of the corresponding L$_n$Dy(OH) complex. Therefore, the dihydrate complex Dy(tta)$_3$(H$_2$O)$_2$ was gradually heated in ultra-high vacuum and the formation of volatile species followed by mass spectrometry (electron ionization, direct inlet). It was found that both water ligands were cleaved off and evaporation of the anhydrous Dy(tta)$_3$ complex was detected above 463\,K. No formation of free 1,3-diketone, which would indicate partial ligand hydrolysis during the heat-up process, was observed below that temperature. These studies show that anhydrous Dy(tta)$_3$ can be prepared and evaporated without decomposition. 

These molecules were evaporated from a Knudsen cell at 470\,K onto an atomically clean Au(111) surface held at room temperature in ultra-high vacuum. The sample was then annealed to 385\,K to allow for self-assembly on the surface. 

\subsection{Scanning Tunneling Microscopy} The prepared sample was cooled down and transferred under ultra-high vacuum conditions into a custom-made scanning tunneling microscope (STM) with a working temperature of 4.8\,K. All STM images were recorded in constant-current mode with a Au-coated tungsten tip. Differential conductance spectra were acquired with fixed tip-sample distance or activated feedback-loop as indicated in the respective figure captions, using a lock-in amplifier. 

\subsection{X-ray Spectroscopy} X-ray absorption (XA) spectra were measured at the high-field diffractometer of the beamline UE46-PGM1 at BESSY II, using $p$-linearly or circularly polarized X rays. O-$K$ XA spectra were measured with the third harmonic and Dy- and Gd-$M_{4,5}$ XA spectra with the fifth harmonic of the undulator. Linear and circular degrees of polarization were about 99\% and 85\%, respectively. The energy resolution was set to approximately 160~meV, resulting in a photon flux density of about $10^{10}$ photons$\cdot$s$^{-1}$mm$^{-2}$ with a spot size of about 1~mm$^2$. No time-dependent spectral changes at the O-$K$, Dy-$M_{4,5}$, and Gd-$M_{4,5}$ edges have been observed on  the timescale of the experiment. Furthermore, no time dependence of the magnetic behavior was  observed. We, thus, exclude X-ray-induced degradation of the molecules for the presented results. The XA signal was recorded in total electron yield mode measuring the drain current of the sample as a function of photon energy.  This signal was normalized to the total electron yield of a gold grid upstream to the experiment and to the spectrum of a clean Au(111) substrate without adsorbed molecules.  For the X-ray magnetic circular dichroism (XMCD) measurements, an external magnetic field was applied parallel to the photon propagation direction. All X-ray natural linear dichroism (XNLD) spectra were measured in a small  magnetic field of 20\,mT applied parallel to the $k$ vector of the X-rays to ensure efficient extraction of the secondary electrons at a temperature of 4.5\,K. All XA measurements for both, Dy(tta)$_3$ and Gd(tta)$_3$, were performed at the same coverage of 0.2\,ML.

\section{\texorpdfstring{Dy(tta)\boldmath$_3$ on Au(111)}{Dy(tta)3 on Au(111)}} 

\subsection{Molecular Configuration}

\begin{figure}
\centering
  \includegraphics[width=\columnwidth]{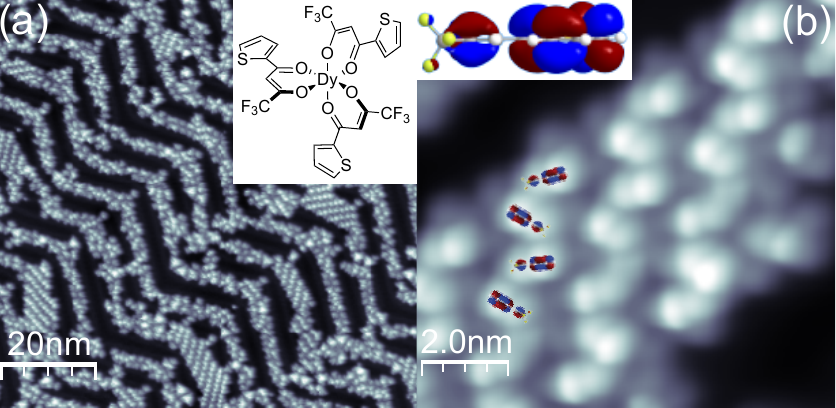}
  \caption{(a) STM image of the Dy(tta)$_3$ complex on Au(111). The islands of densely packed Dy(tta)$_3$ complexes align along the herringbone reconstruction ($V=0.5$\,V, $I=75$\,pA). Inset: Chemical structure of Dy(tta)$_3$. (b) Close-up STM image of the molecular islands. A regular zigzag alignment of the molecules within these islands can be observed. A smaller and larger lobe separated by a nodal plane can be identified ($V=0.3$\,V, $I= 50$\,pA). Inset: Top view of the LUMO of one negatively charged tta ligand, calculated with DFT (basis set B3LYP/6-31+G$^*$).}
  \label{fig1}
\end{figure}

\begin{figure}
  \includegraphics[width=0.48\textwidth]{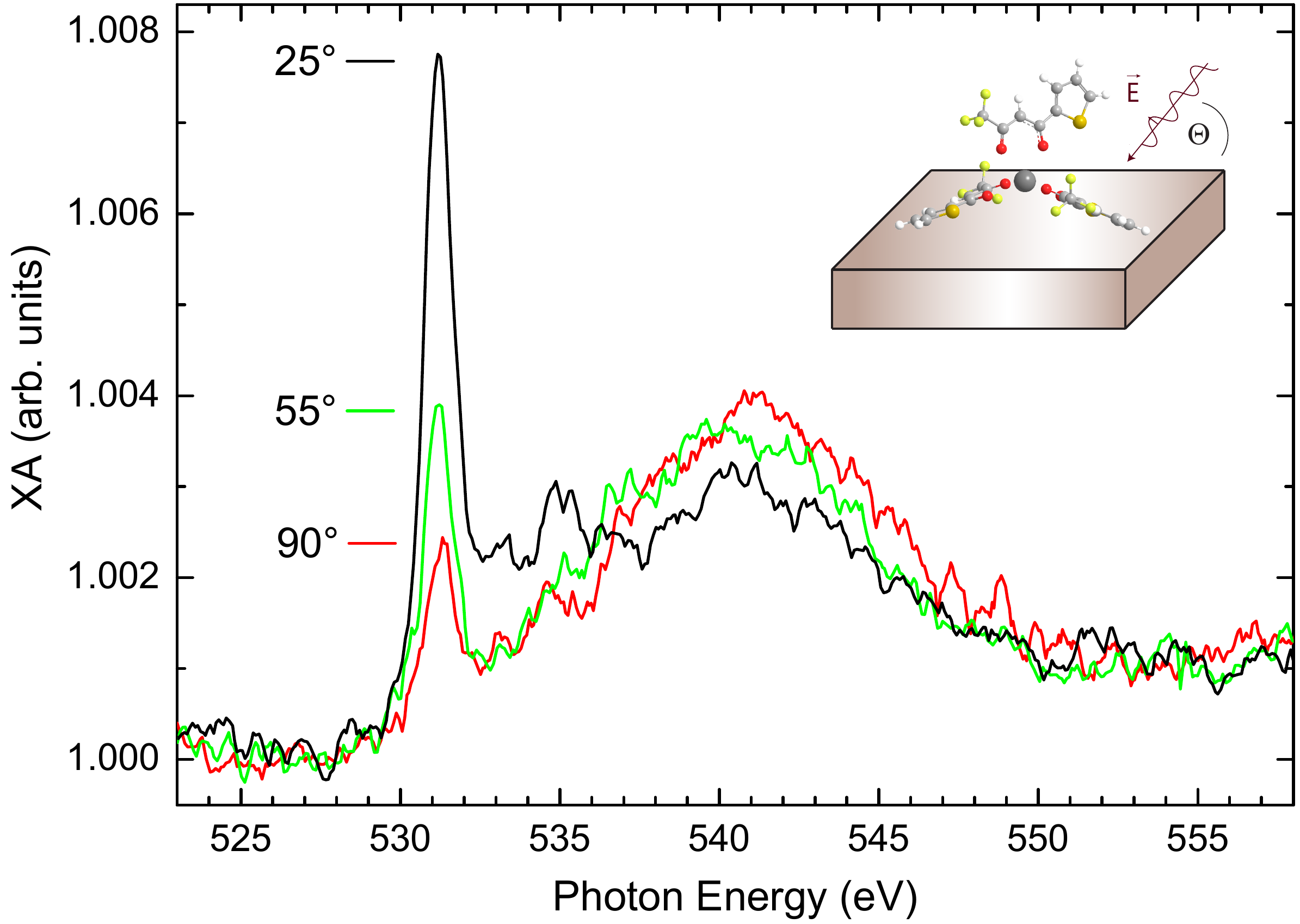}
  \caption{X-ray absorption spectra of the O $K$ edge of 0.2\,ML Dy(tta)$_3$ on Au(111), recorded with angles of 25$^{\circ}$, 55$^{\circ}$, and 90$^{\circ}$ between the polarization vector of the linearly polarized X rays and the surface normal. Inset: Sketch of the adsorption geometry with direction of X-ray beam and $E$-field vector.}
 \label{fig2}
\end{figure}

Deposition of the Dy(tta)$_3$ molecules at room temperature onto a clean Au(111) surface under ultra-high vacuum conditions and post-annealing to 385\,K leads to densely packed molecular arrangements, which align along the herringbone reconstruction of the substrate (Fig.\,\ref{fig1}). A close-up view of the STM images reveals features of uniform appearance. Each unit consists of two bright oval-shaped protrusions with slightly different apparent height and of background protrusions, which are less well-defined (Fig.\,\ref{fig1}(b)). The size of these units matches well with the molecular size and their uniformity reveals that the molecules are intact on the surface. Comparison of the LUMO shape of the singly-charged tta ligand (superimposed on the STM image) with the oval protrusions suggests that one tta moiety is standing upright with respect to the surface. The other two ligands are seen as the lower protrusions in the STM images, partially located underneath the upper ligand (sketch in Fig.\,\ref{fig2}, inset). 

\begin{figure}
  \includegraphics[width=0.48\textwidth]{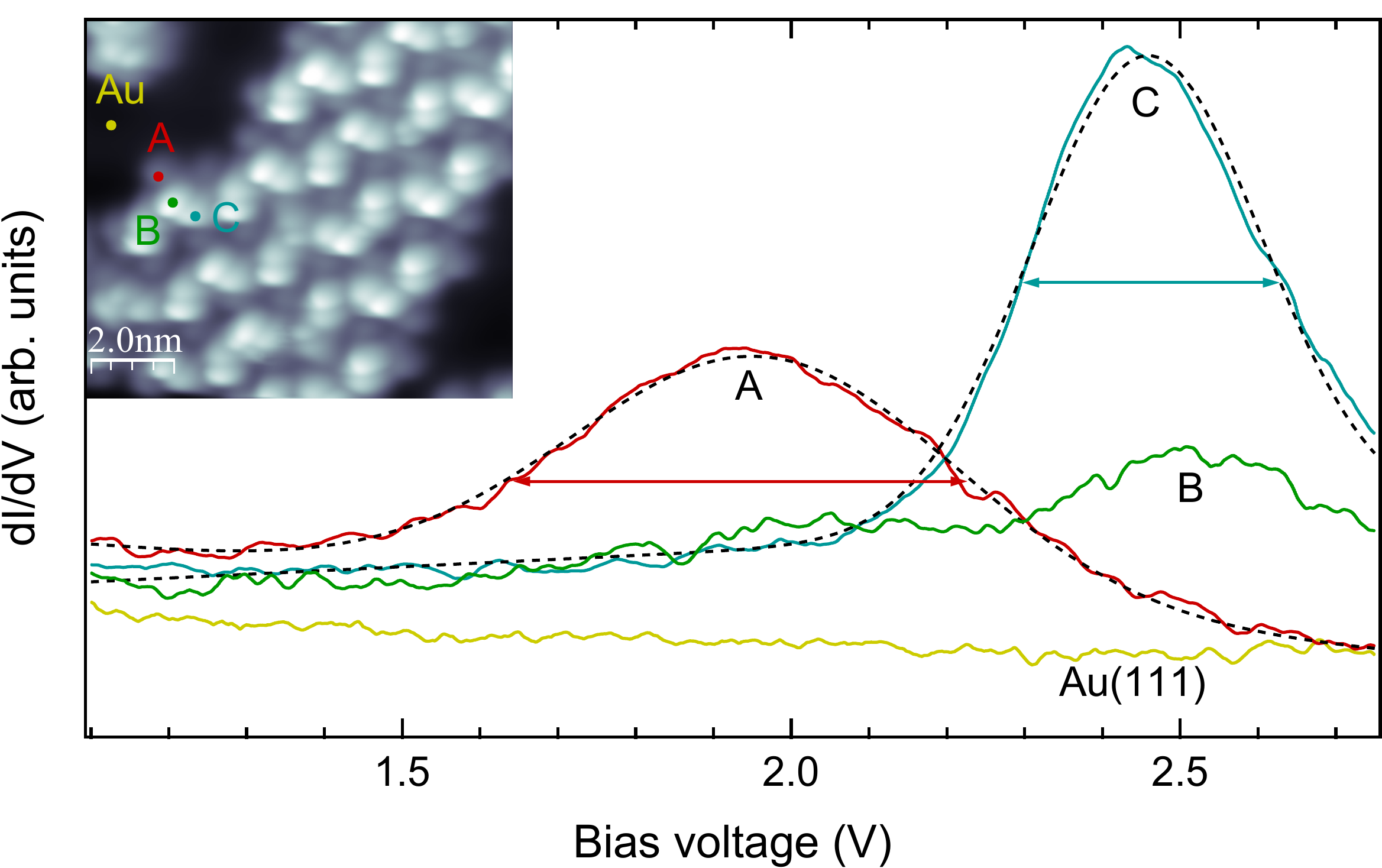}
  \caption{Differential conductance spectra recorded in constant-current mode at different sites of the Dy(tta)$_3$ molecules as color-coded onto the STM image. Spectra A, C, and on gold were recorded at a feedback current of 100\,pA ($V_\mathrm{mod}=5$\,mV), the spectrum at site B at $I=59\,$pA ($V_\mathrm{mod}=15$\,mV). The spectrum recorded on the lower ligand (A, red) shows a broad resonance at 1950\,meV, the one at the edge of the upper ligand (C, blue) shows a sharper resonance shifted to higher energy (2460\,meV). The spectrum at the center (B, green) of the molecule shows a double-peak structure due to both contributions. Red and blue arrows indicate the full width at half maximum (FWHM) of the fitted Gaussian line shapes (dashed lines). The reference spectrum on Au (yellow) is flat.}
 \label{fig4}
\end{figure}

This adsorption scenario leads to the assumption that the complex changes coordination geometry upon binding to the surface. A more reliable determination of the orientation of the ligands on the surface can be obtained by near-edge X-ray absorption fine structure (NEXAFS). 
Figure\,\ref{fig2} shows angle-dependent spectra taken at the oxygen $K$ edge, which represent transitions from the O $1s$ core levels to unoccupied molecular states. All spectra exhibit a pronounced $\pi^*$ resonance at 531.2 eV photon energy with its intensity being highest when the incidence angle of the X rays is strongly grazing. 
This situation corresponds to the polarization vector of the exciting X rays being closest to the surface normal (see sketch in Fig.\,\ref{fig2}, inset). Therefore, the $\pi$ electronic systems around the oxygen atoms, and thus the C--O bonds, must on average exhibit an orientation more parallel to the surface.
Assuming random azimuthal orientations of the molecules as also seen in STM images, we quantitatively evaluate the NEXAFS spectra. The measured angle dependence matches a scenario in which two of the ligands are fully parallel to the surface, while the third one is standing upright with its plane parallel to the surface normal, consistent with the STM results.\\

\subsection{Electronic Structure}
\label{elstruc}

\begin{figure}
\center
 \includegraphics[width=0.48\textwidth]{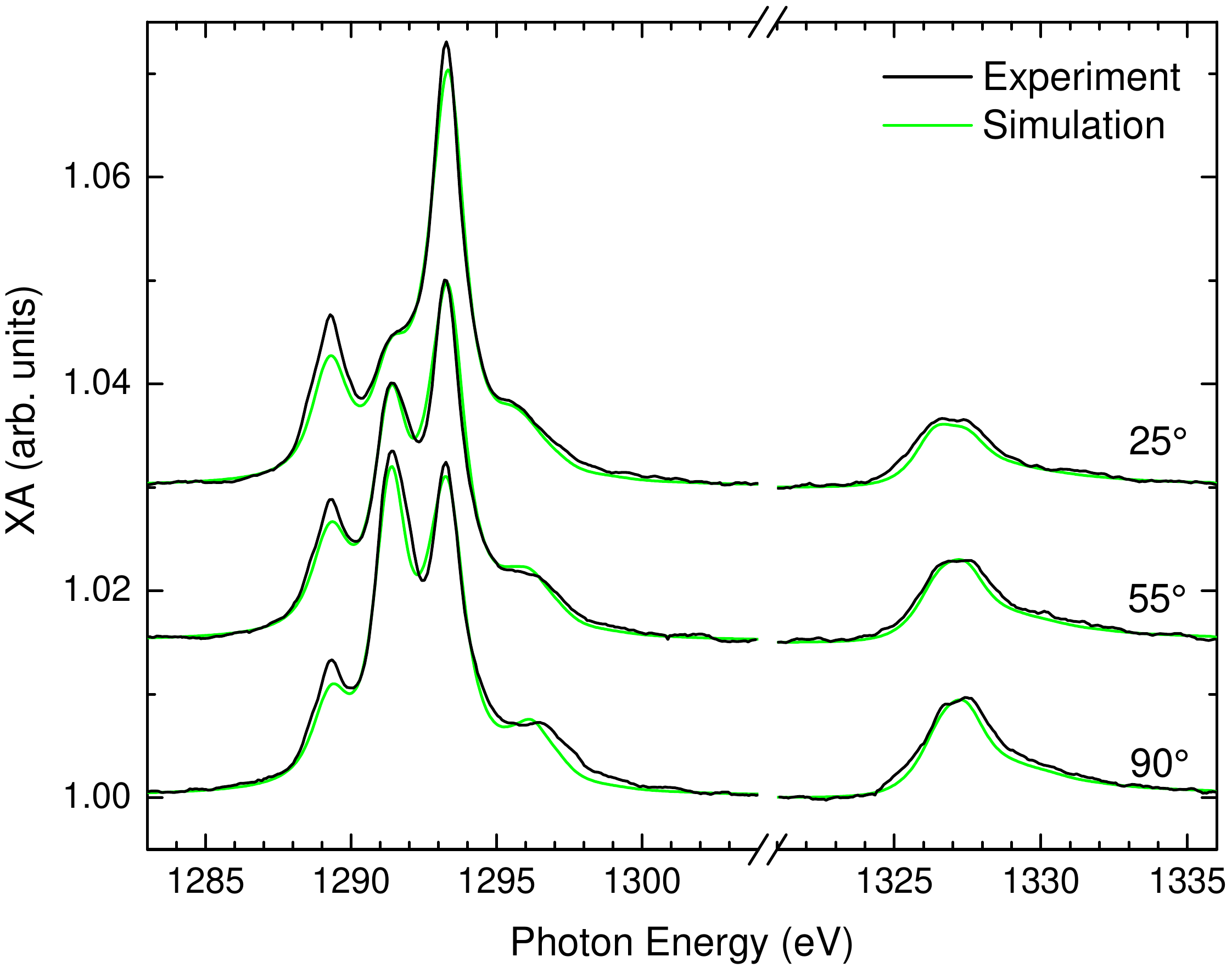}
  \caption{XA spectra (black lines) of 0.2\,ML Dy(tta)$_3$ on Au(111), recorded at the Dy-$M_{4,5}$ edges with angles of $25^\circ$, $55^\circ$, and $90^\circ$ between the $E$ vector of the linearly polarized X rays and the surface normal at a temperature of 4.5\,K. The characteristic triple-peak structure evidences transitions from the filled $3d$ shell to the open $4f$ shell with $\Delta J=0,\pm1$. Green lines are simulated spectra obtained from multiplet calculations. The spectra are offset for clarity.}
  \label{Dy-xnld}
\end{figure}

The geometry of the adsorbed molecules is also reflected in the electronic structure.
We recorded differential conductance spectra (d$I$/d$V$) with sub-molecular resolution (Fig.\,\ref{fig4}). Spectra taken on the upper ligands (location C) exhibit a resonance at 2460\,meV  with a full width at half maximum (FWHM) of 350\,meV. Spectra on the lower ligand (location A) show a peak that is energetically down-shifted by about 500\,meV and significantly broader (FWHM\,=\,580\,meV). This behavior evidences a stronger hybridization with the substrate, leading to energy-level broadening and downshifting of the resonance due to stronger screening of the tunneling electrons~\cite{TorrenteJPCM08}. When tunneling through both types of ligands, \textit{i.e.}, through the center of the molecule (location B), both peaks can be detected simultaneously. Hence, the spatially resolved spectra corroborate the picture of one upright-oriented ligand, which is hardly affected by the underlying substrate, and two ligands, which are almost flat on the gold surface.

To investigate how the electronic structure of the Dy ions reacts to the electrostatic field imposed by the tta ligands and the surface, we have carried out X-ray absorption (XA) measurements at the Dy $M_{4,5}$ edges using linearly and circularly polarized X rays.
Figure\,\ref{Dy-xnld} shows the XA spectra recorded at different X-ray incidence angles for linearly polarized X rays.
The peaks correspond to transitions from the filled $3d$ shell to the open $4f$ shell with the lower energy part deriving from the $3d_{5/2}$ and the higher energy transitions deriving from the $3d_{3/2}$ states, respectively.  
The characteristic triplet structure at the $M_5$ edge stems from transitions with $\Delta J= 0,\pm1$.
The line shape agrees with a $3+$ oxidation state \cite{GoedkoopPRB88}, as expected for the molecule with three monovalent tta-ligands. Furthermore, they evidence the predominantly atomic character of the Dy, \textit{i.e.}, that of the free ion, which shows that the orbital moment is hardly affected by the surrounding ligands.

\begin{figure}
\center
  \includegraphics[width=0.48\textwidth]{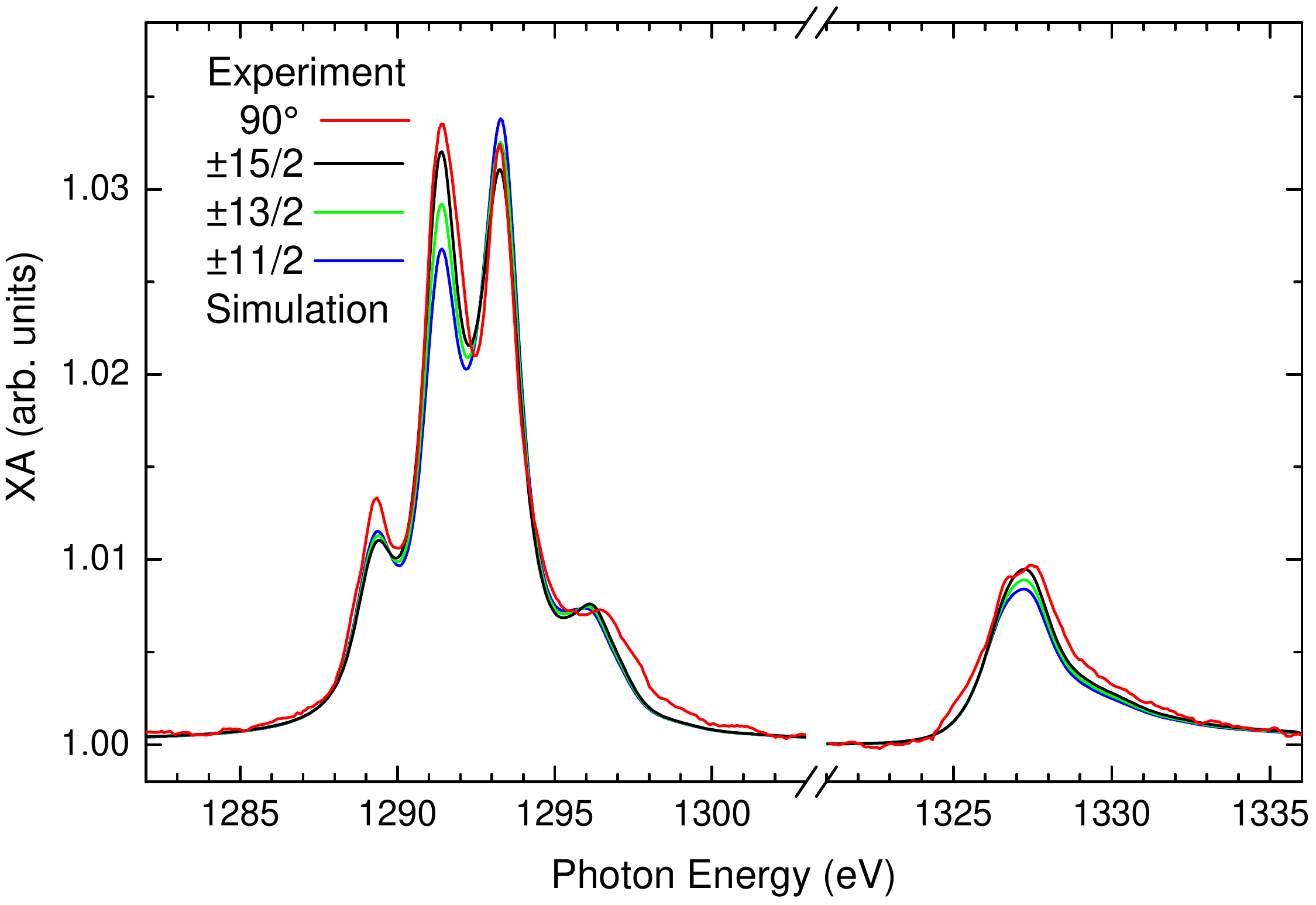}
  \caption{Calculated Dy $M_{4,5}$ XA spectra for three initial-state doublets and experimental spectrum measured with linearly polarized X rays and the $E$ vector parallel to the surface.}
  \label{mj}
\end{figure}

The intensity distribution within the triplet peak structure at the Dy $M_5$ edge depends on the $M_J$ ground state of the Kramers-split eigenstates. To identify the ground state, spectra have been fitted using multiplet theory~\cite{GoedkoopPRB88,Cowan81}, as described in Appendix\,\ref{calcu}.
Calculated spectra for three initial-state doublets with maximum angular momentum projection are compared in Fig.\,\ref{mj} to the experimental Dy $M_{4,5}$ XA spectrum measured with linearly polarized X rays and the $E$ vector parallel to the surface at a temperature of $4.5$\,K.  The resonance at $1291.4$\,eV corresponding to $\Delta J = 0$ is decreasing with decreasing absolute value of the magnetic quantum number (see Appendix \ref{kramer}). Clearly, the experimental spectrum corresponds to an $M_J=\pm 15/2$ ground state. 
In the well-known Dy-bis(phthalocyaninato) complexes \cite{ishikawa03} with a cylindrical charge density distribution, in contrast, such a ground state is disfavored and an $M_J=\pm 13/2$ ground state is observed \cite{rinehart11}.

\begin{figure}
\center
  \includegraphics[width=0.48\textwidth]{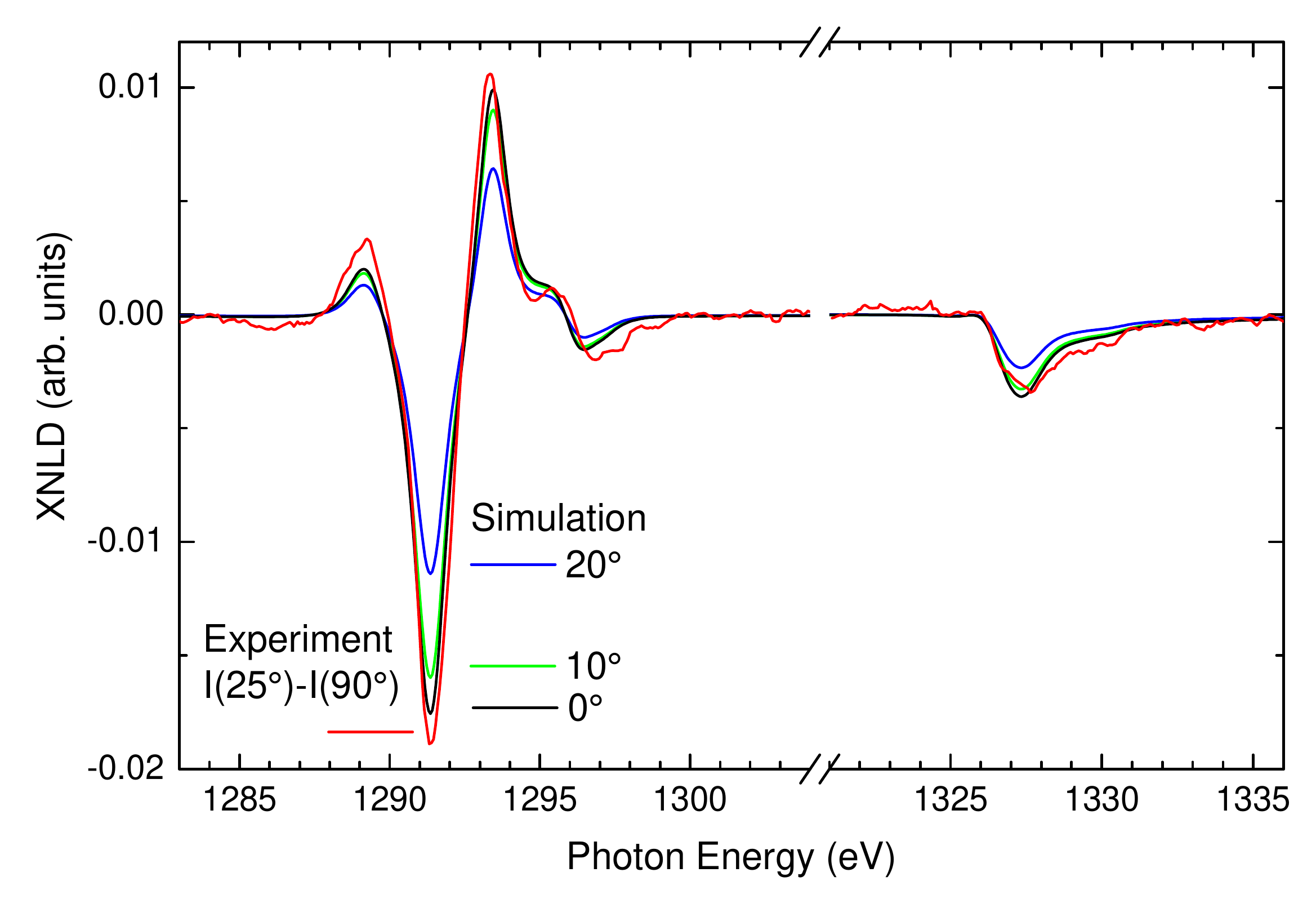}
  \caption{Dy $M_{4,5}$ XNLD difference spectrum between the spectra shown in Fig.\,\ref{Dy-xnld}, measured with linearly polarized X rays with $25^\circ$ and $90^\circ$ between the polarization vector of the X rays and the surface normal. This spectrum is compared to simulated difference spectra for three tilting angles of $0^{\circ}$, $10^{\circ}$, and $20^{\circ}$ of the symmetry axis of the $f$ orbitals with respect to the surface.} 
  \label{Dyxnld}
\end{figure}

The pronounced variation of relative intensities of the absorption peaks with the incidence angle of the X rays indicates a distinct orientation of the Dy $4f$ orbitals on the surface. 
To determine the orientation of the orbitals, the difference spectrum between spectra measured with $25^\circ$ and $90^\circ$ incidence angle is compared to simulated spectra for three different orientations of the symmetry axis of the orbitals with respect to the surface (Fig.\,\ref{Dyxnld}), assuming a uniaxial anisotropy.  The simulated difference spectra show a strong influence of the orientation of the symmetry axis with respect to the surface. The largest difference is expected for the symmetry axis parallel to the surface. The experimental difference spectrum shows maximum difference, thus reflecting an orientation of the symmetry axis parallel to the surface.

\subsection{Magnetic Properties}

\begin{figure}
  \includegraphics[width=0.48\textwidth]{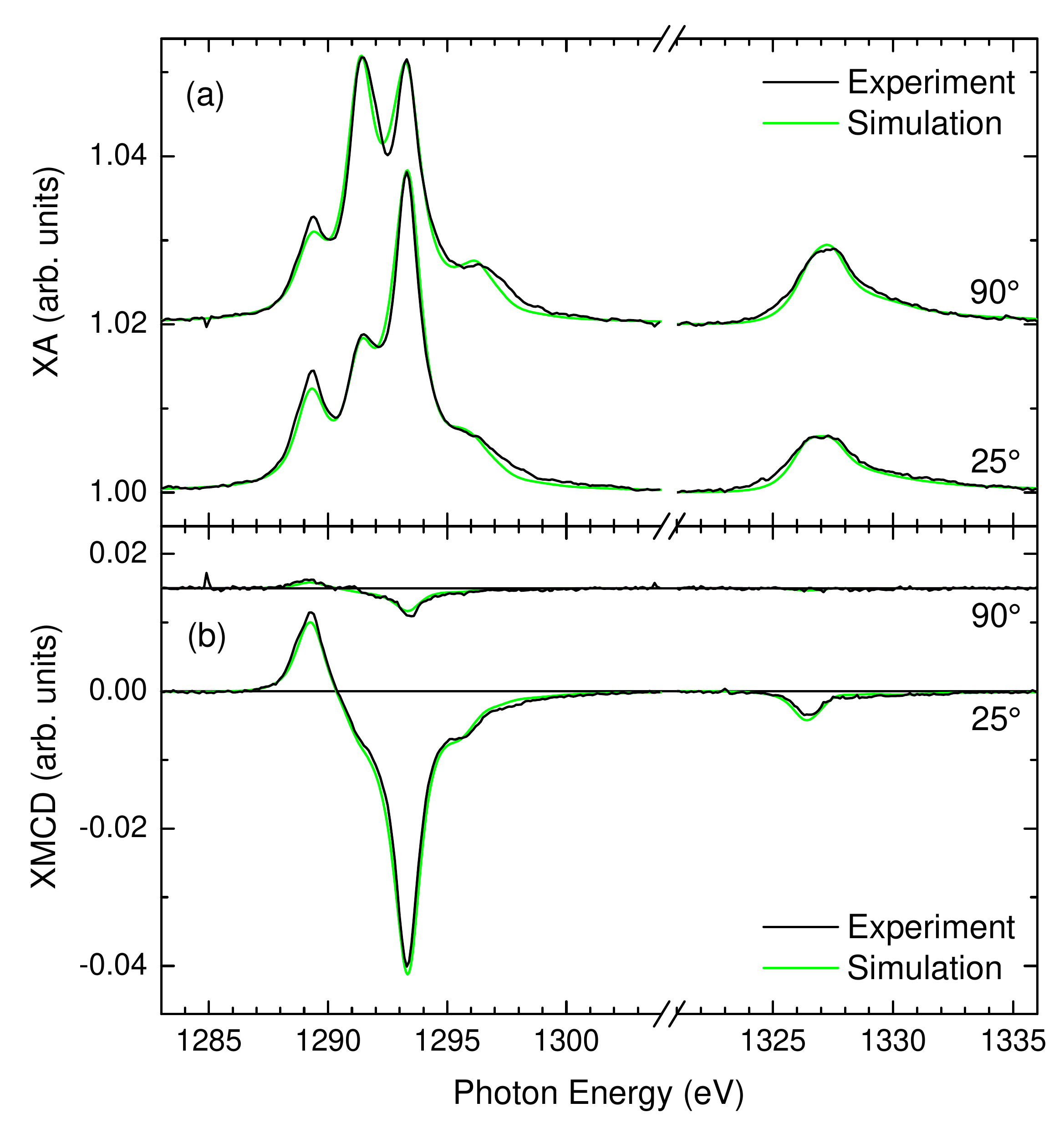}
  \caption{(a) Dy-$M_{4,5}$ XA spectra (black lines) recorded under 90$^\circ$ and 25$^\circ$ incidence angles in an applied magnetic field of 6\,T parallel to the $k$ vector of the circularly polarized X rays at a temperature of 4.5\,K. The spectra are offset for clarity. (b) The XMCD signal reflects a sizable magnetic moment of the Dy core for $25^{\circ}$ incidence. Green lines are simulated spectra obtained from multiplet calculations. 
  }
 \label{Dy-xmcd}
\end{figure}

\begin{figure}
  \begin{tabular}{c}
    \includegraphics[width=0.48\textwidth]{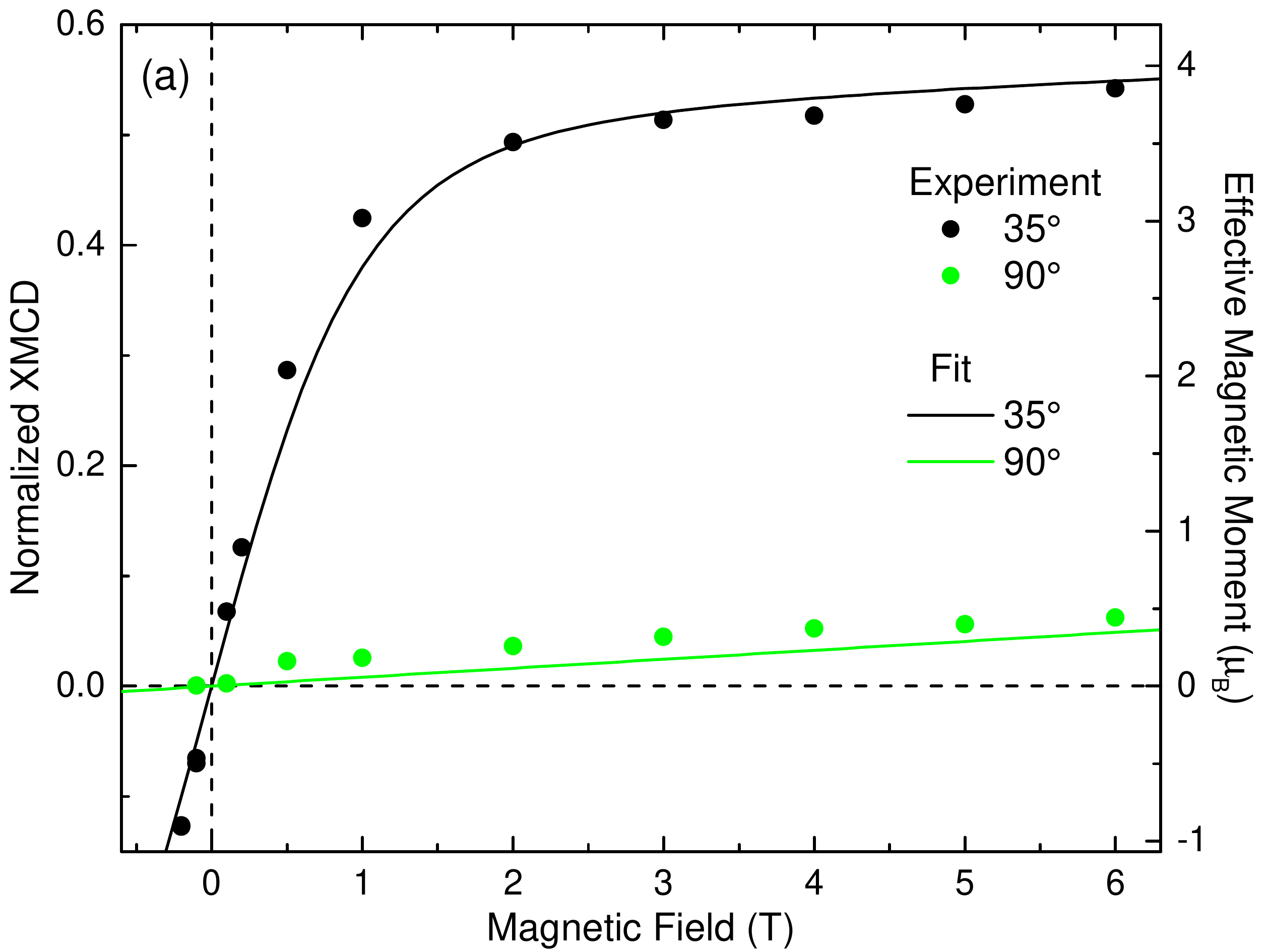}\\
    \includegraphics[width=0.45\textwidth]{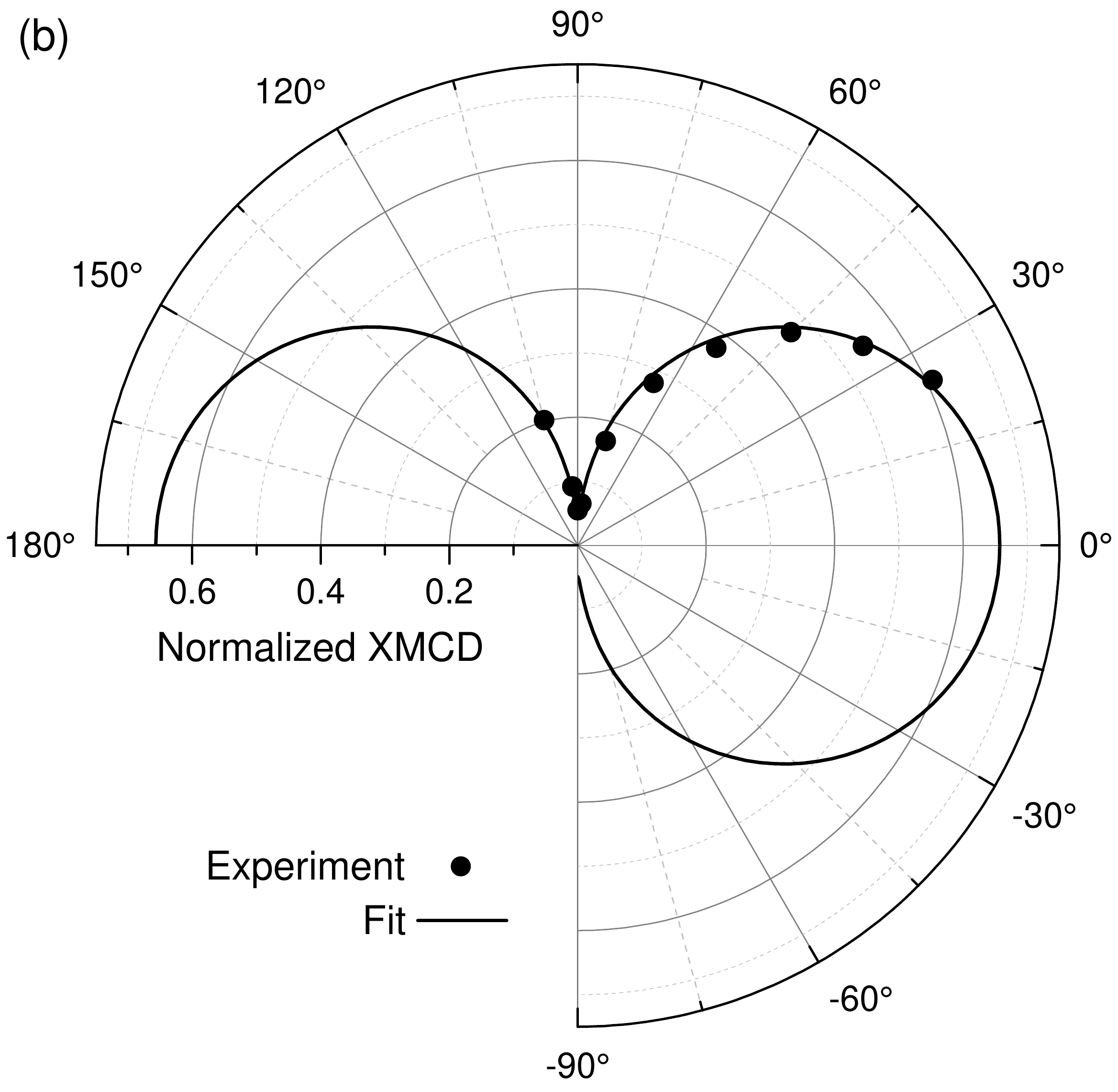}
  \end{tabular}  
  \caption{(a) Magnetic field dependence of the integrated Dy-$M_5$ XMCD signal at $35^\circ$ and $90^\circ$ incidence angles. The right y axis displays the experimentally determined effective magnetic moment derived from a sum-rule analysis (see Appendix\,\ref{effmom}). (b) Angle dependence of the integrated Dy-$M_5$ XMCD in an applied magnetic field of 6\,T. Measurements were carried out at a temperature of 4.5\,K with magnetic field and X-ray beam along the indicated angles.  The lines are a fit of the model described in the text. The XMCD signal has been normalized to its saturated value.}
   \label{mfit}
\end{figure}

Concomitant with the distinct orientation of the orbitals, one may expect a strong magnetic anisotropy in the case of a partially filled $4f$ shell, which is the key ingredient of molecular magnets. XMCD signals are the difference between two spectra recorded with opposite helicities of circularly polarized X rays and are proportional to the magnetization projected onto the $k$ vector of the X rays.  Dy $M_{4,5}$ XA spectra for circularly polarized X rays and the corresponding XMCD difference curves taken in a magnetic field of 6\,T and a temperature of 4.5 K are presented in Fig.\,\ref{Dy-xmcd}(a) and (b), respectively.  The XMCD signal at $25^{\circ}$ grazing incidence is 8.7 times higher compared to the one in the perpendicular direction.  This evidences an easy axis of magnetization  parallel to the surface.  Such alignment is different to lanthanide bis(phthalocyaninato) molecules on surfaces, where the easy axis of magnetization is perpendicular to the surface \cite{Klar14,Stepanow10,Wackerlin16}.
The strong angle dependence of the XMCD signal hints at a large magnetic anisotropy due to the distinct distortion in the ligand field. 
To quantify this magnetic anisotropy we recorded the XMCD signal as a function of the magnitude and the direction of the external magnetic field.  The integrated Dy $M_5$ XMCD signal at $35^\circ$ and $90^\circ$ incidence angles is shown in Fig.\,\ref{mfit}(a). The magnetization curve at $35^\circ$ shows a steep increase and seems to saturate already at about 2\,T. The XMCD signal in the vertical direction is very small and does not reach saturation up to 6\,T. The integrated Dy $M_5$ XMCD as a function of angle between the magnetic field and the surface is shown in Fig.\,\ref{mfit}(b) for $B=6\,$T and $T=4.5\,$K. The XMCD is maximum at grazing directions and minimum when the magnetic field is applied normal to the surface.
Unlike in lanthanide bis(phthalocyaninato) complexes where an unpaired electron is delocalized on the phthalocyanine ligands \cite{Komeda11}, each of the three tta ligands in the free Dy(tta)$_3$ complex is singly, negatively charged but has an even number of electrons. Since the Dy ion and the ligands are in their stable oxidation states a scenario in which additional unpaired spins on the ligands are created upon adsorption onto Au(111) is highly unlikely.
Due to the shielded nature of the $4f$ electrons, the angular momentum $L$ is not expected to be quenched by the ligand field and we can treat spin--orbit coupling in the Russel-Saunders ($LS$) scheme with the lowest multiplet given by Hund's rules as $J=15/2$ for the nine $4f$ electrons of a Dy$^{3+}$ ion \cite{nolting}.
In its simplest form the ligand-field-induced anisotropy can be expressed as a uniaxial anisotropy with an axial zero-field splitting parameter $D$.
The effect of the magnetic field is given by the Zeeman term. We thus model our system with the following Hamiltonian:
\begin{equation} 
H=\,\mu_B g_J \vec{B}\cdot\vec{J}+D J_{z}^2\,,
\end{equation}
with $g_J=4/3$ the Land{\'e} g factor.
To calculate the field-dependent integrated XMCD, we start from the thermally populated eigenstates of the Hamiltonian and sum over all allowed X-ray-induced transitions. 
Averaging over different azimuthal incidence directions and orientations of the magnetic field is carried out to account for the random azimuthal orientations of the Dy ions. 
The fit parameter is the anisotropy parameter $D$ with the orientation of the symmetry axes of the Dy$^{3+}$ ions aligned parallel to the surface, as determined from the Dy $M_{4,5}$ XNLD spectrum shown in Fig.\,\ref{Dyxnld}.  This parameter is also used for a simultaneous fit of all XA spectra (green lines in Figs.\,\ref{Dy-xnld} and \ref{Dy-xmcd}).  

\begin{figure}
  \includegraphics[width=0.46\textwidth]{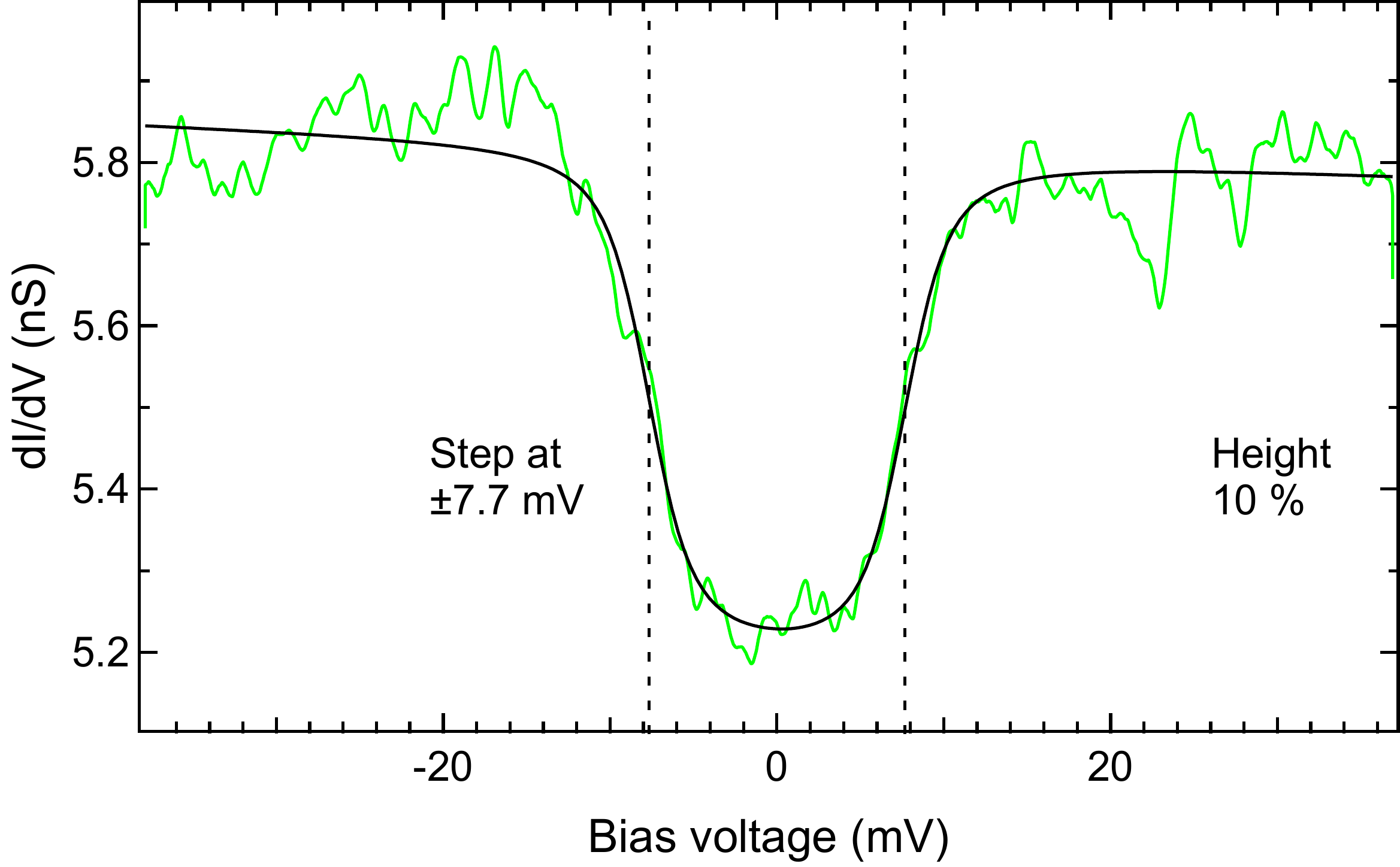}
  \caption{STS spectrum revealing inelastic excitations on Dy(tta)$_3$. The spectrum was recorded in constant-height mode ($V_\mathrm{mod}=1\,$mV) with the feedback opened at $I=0.3\,$nA and $V=60\,$mV at 4.8\,K. Black lines are fits using an arctan-step function to determine the step positions ($\pm$7.7(3)\,mV and heights 10$\%$ of the d$I$/d$V$ amplitude).}
 \label{fig7}
\end{figure}

\begin{figure}
\center
  \includegraphics[width=0.48\textwidth]{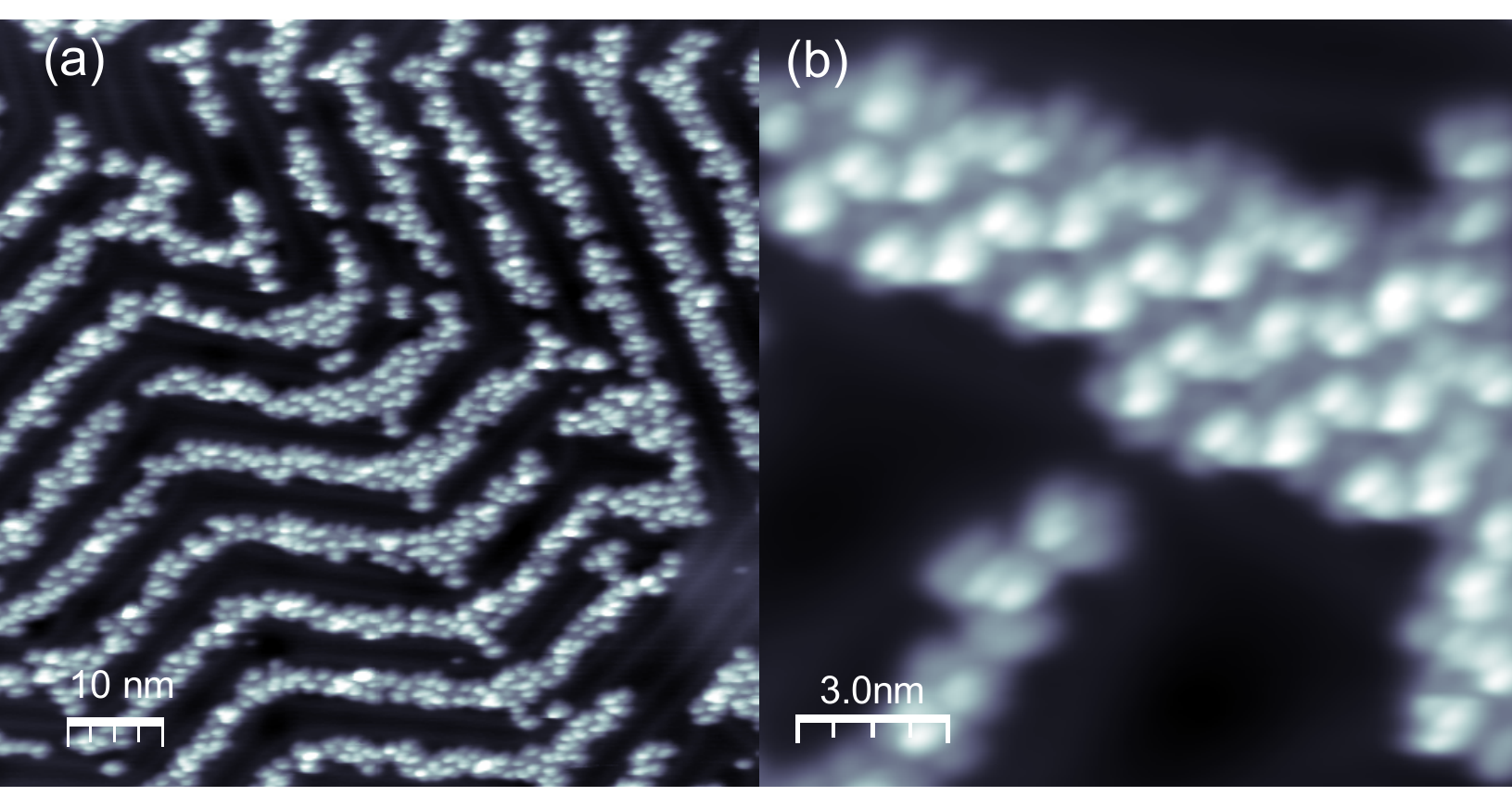}
  \caption{STM images of Gd(tta)$_3$ on Au(111) ($V=0.65\,$V, $I=70\,$pA). They reveal the close resemblance to the molecular arrangement and orientation of Dy(tta)$_3$ on Au(111).} 
  \label{topos}
\end{figure}

\begin{figure}
  \includegraphics[width=0.48\textwidth]{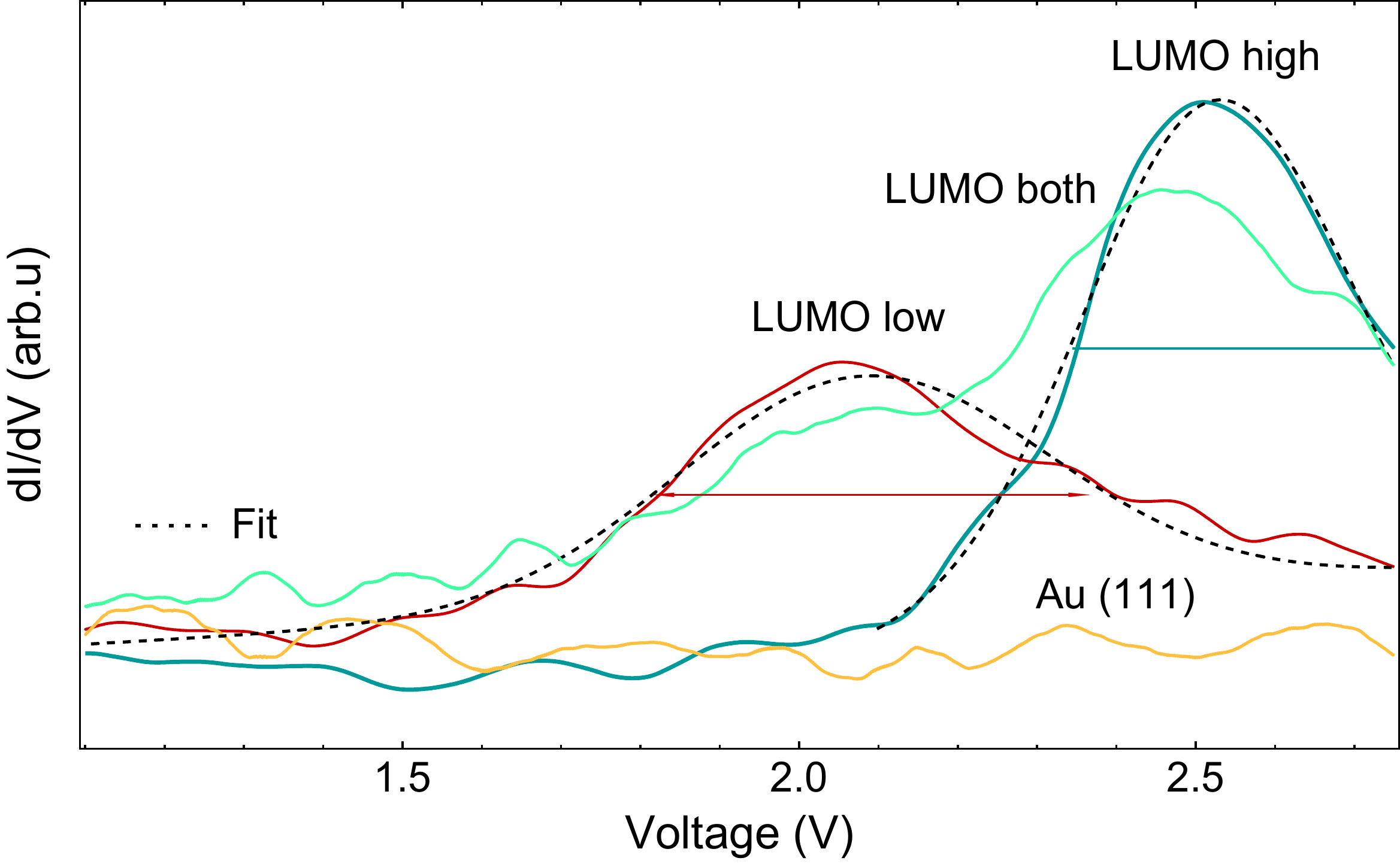}
  \caption{d$I$/d$V$ spectra on Gd(tta)$_3$ on Au(111) recorded in constant-current mode ($I=50\,$pA, $V_\mathrm{mod}=10\,$mV) at 4.8\,K: a peak at 2080\,mV with FWHM of 549\,mV is found on the lower protrusion (red) and a peak at 2523\,mV, FWHM of 389\,mV is found on the higher protrusion (blue). Both peaks appear simultaneously when tunneling through both ligand types in the center of the molecule (green). The peak positions vary by $\approx$\,100\,mV depending on the arrangement of the neighbouring molecules.}
  \label{LUMOs}
\end{figure}

\begin{figure*}
\begin{tabular}{cc}
  \includegraphics[width=0.48\textwidth]{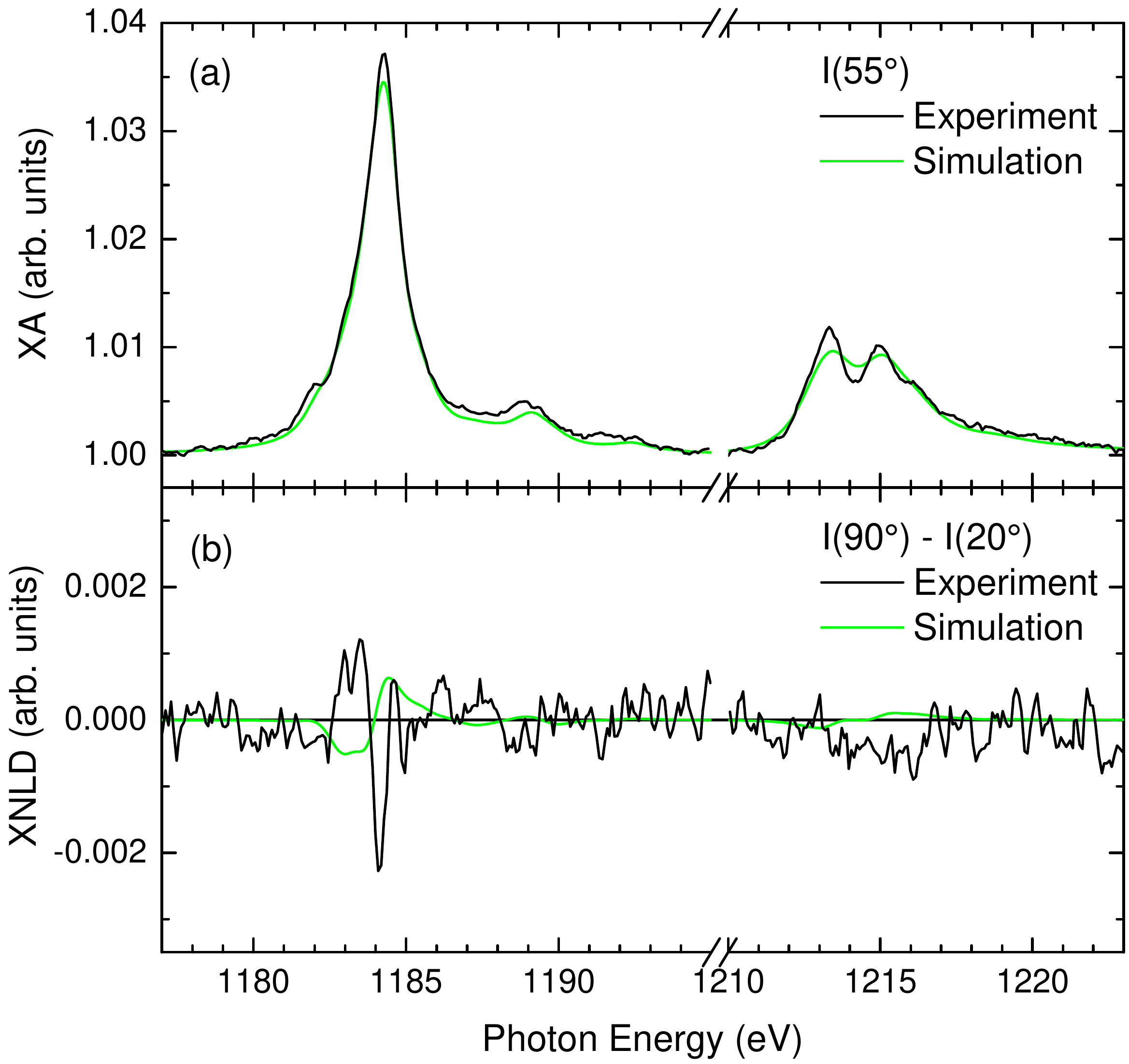}&
  \includegraphics[width=0.48\textwidth]{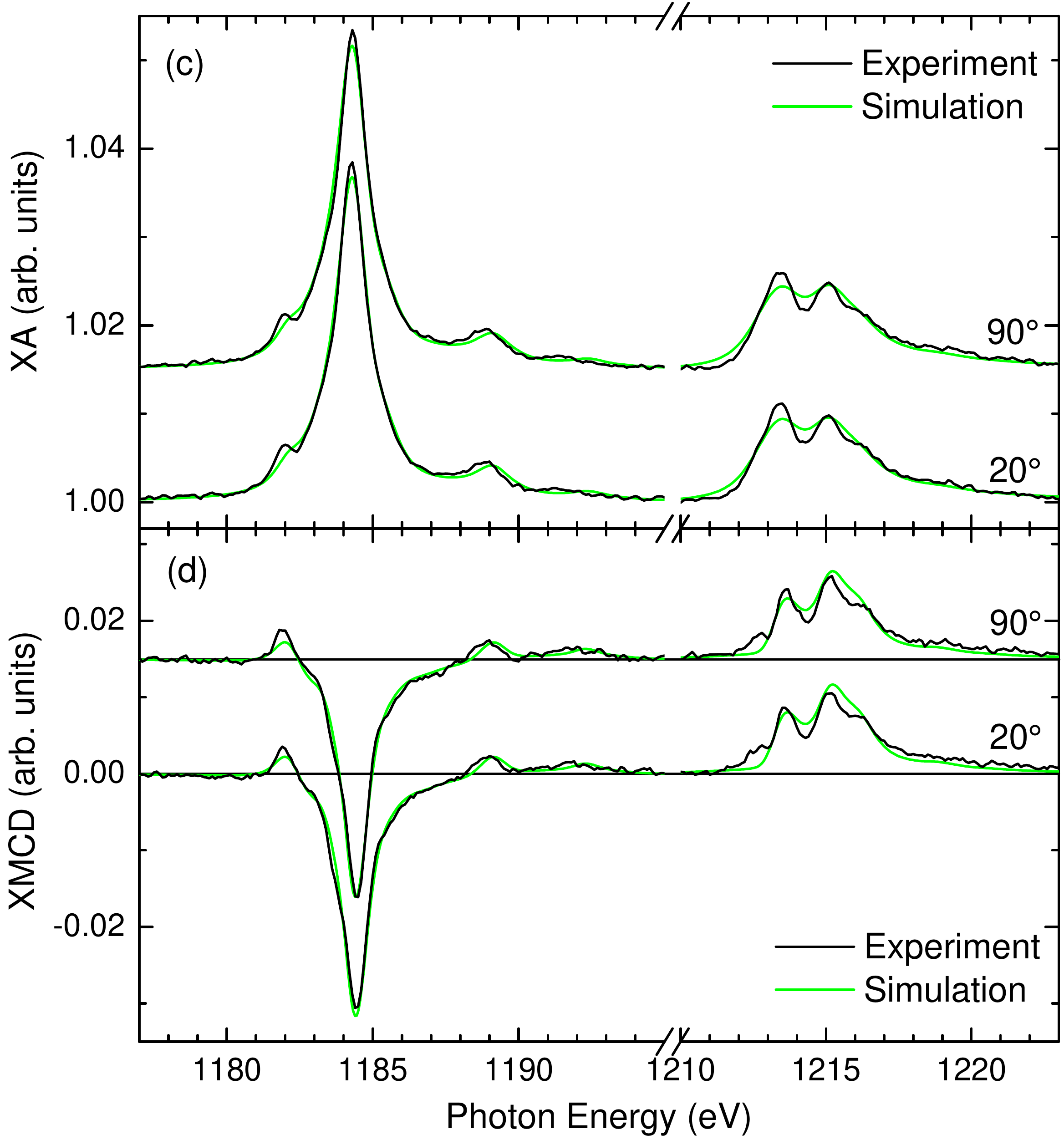}
\end{tabular}
  \caption{
  Angle-dependent Gd $M_{4,5}$ XA (a), (c), XNLD (b) and XMCD (d) of 0.2\,ML Gd(tta)$_3$ on Au(111) recorded at $T=4.5\,$K. The spectra shown in panels (a) and (b) are measured in a small magnetic field of 20\,mT with angles of $20^\circ$, $55^\circ$, and $90^\circ$ between the $E$ vector of the linearly polarized X rays and the surface normal. The spectra shown in panels (c) and (d) are measured in a magnetic field of $6\,$T applied parallel to the $k$ vector of the circularly polarized X rays. Simulated spectra fitted to the experimental ones are shown in green.
  }\label{Gd-xas}
\end{figure*}

From the fit (lines in Fig.\,\ref{mfit}) we determine a strong easy-axis anisotropy of the Dy$^{3+}$ magnetization with $D=-0.68(15)\,$meV.
The steep increase at low magnetic field can only be explained with an $M_J=\pm15/2$ ground state. This agrees with the fit of the XA spectra, see Fig.\,\ref{mj}.  Notably, the magnetization at $35^\circ$ only reaches about 54\% of its saturated value at maximum field, due to the random azimuthal orientations of the anisotropy axes of the molecules. The particular orientation of the preferred magnetization parallel to the surface can be rationalized from the adsorption configuration of the molecules and highlights the importance of the low-symmetry ligand field: the anisotropy axis lies in the plane of the two planarly adsorbed ligands, which indicates that the ligands orient the oblate-shaped $4f$-shell electron density of Dy$^{3+}$ and therefore the symmetry axis of the ion. Such orientation in the direction of two of the three ligands has been discussed in DFT calculations, which use an electrostatic approach for the interaction of the gas-phase molecule \cite{Chilton2013}. The calculated magnetization axis lies in the direction of the two opposite-lying charged $\beta$-diketonate ligands.

Assuming the magnetic anisotropy to be uniaxial is a strong simplification, but it is already sufficient to describe the data. We have also tried models containing higher order Stevens operators but the result is an underdetermined situation due to the random azimuthal orientations of the molecules on the surface. An easy-plane anisotropy with an orientation of the symmetry axis of the $f$ orbitals perpendicular to the surface can be excluded as shown in Appendix\,\ref{posD}.

\subsection{Inelastic Tunneling Spectra on Individual Molecules}

The XMCD data unambiguously show that the Dy(tta)$_3$ molecules exhibit a significant magnetic anisotropy with a well-defined energy separation between the ground state and the first excited state. It would be desirable if the magnetic state of an individual molecule could be read out. With the STM tip one can directly address the spin states of individual atoms or molecules and detect their magnetic excitations by inelastic tunneling electrons ~\cite{heinrich04,Hirjibehedin07,Gauyacq2012}. 
With the above derived magnetic ground state of $J=15/2$ and easy-axis anisotropy, \textit{i.e.}, $M_J=\pm15/2$ and $D=-0.68(15)$\,meV, we expect possible inelastic transitions from the $M_J=\pm15/2$ to the $M_J=\pm13/2$ state with $\Delta E=D((15/2)^2-(13/2)^2) =-9.5(2.1)$\,meV. Differential conductance spectra on single molecules indeed exhibit symmetric steps at $\pm7.7(3)$\,meV around the Fermi level with a change of conductance of $\sim$\,10\,\%. This transition could therefore correspond to the inelastic excitation from the $M_J=\pm15/2$ ground state to the $M_J=\pm13/2$ first excited state (Fig.\,\ref{fig7}). Although this is in agreement with the XMCD data, we should unequivocally rule out a different origin, which may be the excitation of molecular vibrations. 
To test this possibility, we investigated isostructural molecules with a Gd$^{3+}$ center. Gd(tta)$_3$ exhibits the same structural and electronic properties as Dy(tta)$_3$. Gd$^{3+}$, however, has a half-filled $4f$ shell. Hence, the total angular momentum is zero and we do not expect a sizable magnetic anisotropy, due to the absence of spin-orbit coupling in the ground state.

\section{\texorpdfstring{Gd(tta)\boldmath$_3$ on Au(111)}{Gd(tta)3 on Au(111)}}

\subsection{\texorpdfstring{Comparison of Adsorption Structure of Dy(tta)\boldmath$_3$ and Gd(tta)\boldmath$_3$ on Au(111)}{Comparison of Adsorption Structure of Dy(tta)3 and Gd(tta)3 on Au(111)}}

The Gd(tta)$_3$ complexes have been evaporated at 470\,K and deposited at room temperature onto the clean Au(111) surface and post-annealed to 385\,K, similar to the preparation with Dy(tta)$_3$. In both cases, we observe close-packed islands, which align with the herringbone reconstruction (compare Fig.\,\ref{topos} and Fig.\,\ref{fig1}). Importantly, the STM images of Gd(tta)$_3$ also reveal a similar appearance of the individual molecules. They consist of a bright oval shape, which we attributed to an upright standing tta ligand, and a lower protrusion associated with two almost flat-lying tta ligands. 
Comparison of the differential conductance spectra reveals the similarity of the frontier molecular orbitals of the Gd- and Dy-complex and thus reflect the equivalent orientation of the molecules on the surface (Fig.\,\ref{LUMOs} and Fig.\,\ref{fig2}).

\subsection{\texorpdfstring{Magnetic Properties of Gd(tta)\boldmath$_3$ on Au(111)}{Magnetic Properties of Gd(tta)3 on Au(111)}}

Angle-dependent Gd $M_{4,5}$ XA spectra for linear polarization are shown in Fig.\,\ref{Gd-xas}(a),(b). The shape of the spectrum recorded at the magic angle (55$^\circ$) is typical for Gd in its 3+ oxidation state. The $4f$ shell is half filled with a spin of $S=7/2$ and a vanishing angular momentum. No significant angle-dependent variation of the spectra is observed, as can be seen from the vanishing XNLD spectrum (Fig.\,\ref{Gd-xas}(b)), given by the difference between the spectra recorded at 90$^\circ$ and 25$^\circ$ X-ray incidence.  Gd $M_{4,5}$ XA and XMCD spectra recorded at $B=6\,$T and $T=4.5\,$K with circular polarization are shown in Fig.\,\ref{Gd-xas}(c) and (d)) for 90$^\circ$ and 20$^\circ$ X-ray incidence. Again, no significant variation of the spectra with the incidence angle is observed. This is also in line with the vanishing angular momentum. All spectra shown in Fig.\,\ref{Gd-xas} are fitted simultaneously with spectra (green lines) obtained from multiplet calculations, see Appendix~\ref{calcu}.

\begin{figure}
\center
  \includegraphics[width=0.48\textwidth]{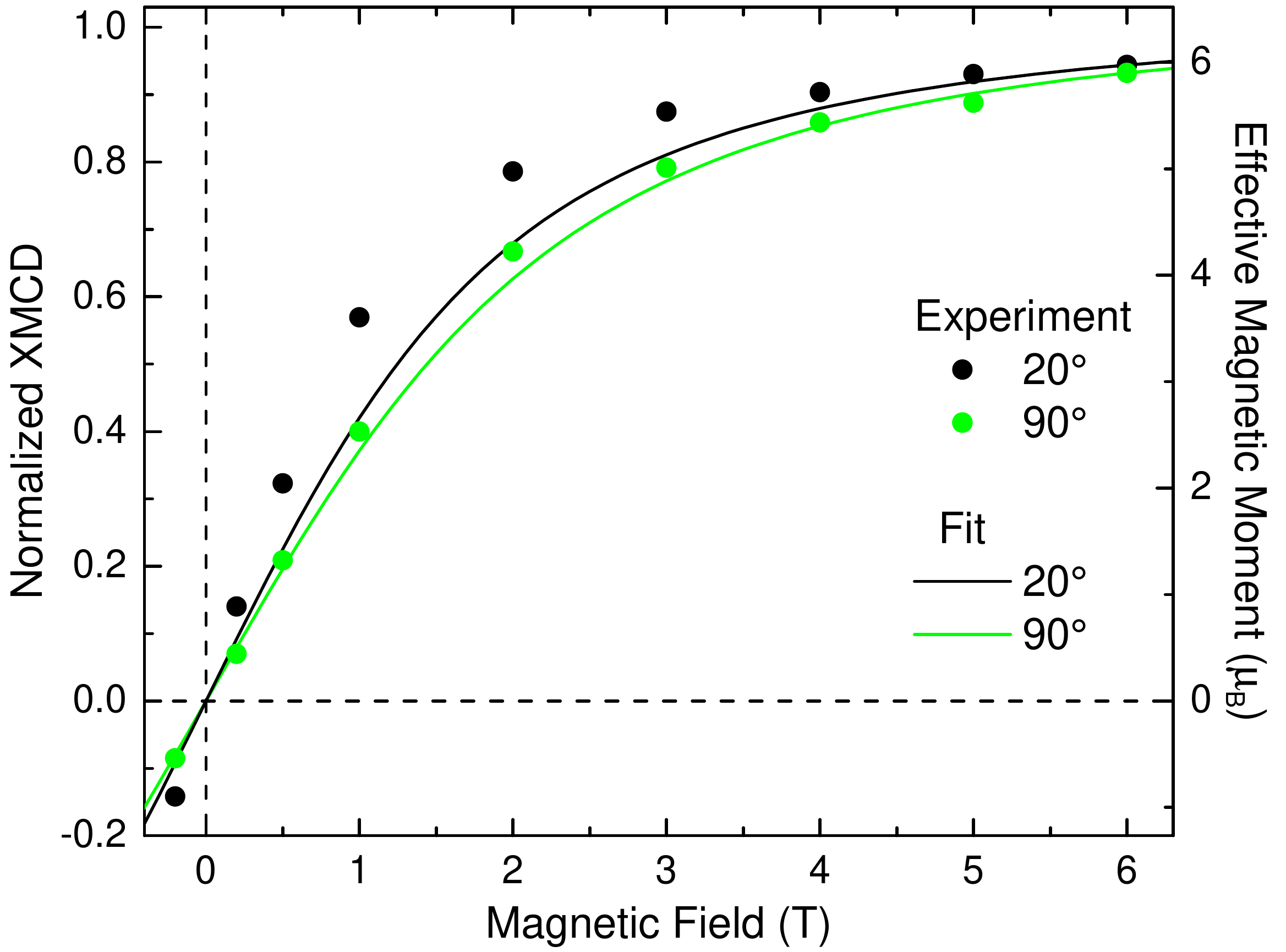}
  \caption{Integrated Gd $M_5$ XMCD signals recorded at 20$^\circ$ (black symbols) and $90^\circ$ (green symbols) incidence angles as a function of external magnetic field at a temperature of $4.5$~K. The right y axis displays the experimentally determined effective magnetic moment derived from a sum-rule analysis (see Appendix\,\ref{effmom}). Lines are a fit to the experimental data of the model described in the main text. The XMCD signal has been normalized to its saturated value.}
  \label{mfitGd}
\end{figure}

In Fig.\,\ref{mfitGd}, the integrated Gd $M_5$ XMCD signal is plotted as a function of magnetic field along the $k$ vector of the X rays for 20$^\circ$ and 90$^\circ$ incidence angles. The XMCD signal at 90$^\circ$ incidence is only slightly smaller than the one at 20$^\circ$, indicating a small magnetic anisotropy. Due to the vanishing angular momentum, such an anisotropy cannot be explained by electrons occupying pure $4f$ states. Presumably, it is a consequence of a slight hybridization between the $4f$ and $5d$ states. This leads to a small magnetic anisotropy that can be described by equation~(1). The difference in magnetization for 20$^\circ$ and 90$^\circ$ incidence can be matched by many combinations of anisotropy parameters and orientations of the anisotropy axes. The theoretical integrated XMCD curves shown in Fig.\,\ref{mfitGd} are a fit to the experimental data assuming anisotropy axes parallel to the surface. The best fit is obtained for $D=-0.02\,$meV. In this situation, the first excited state is only $0.3\,$meV higher in energy than the ground state.

\subsection{Inelastic Tunneling Spectra on Individual Molecules}
\begin{figure}
  \includegraphics[width=0.48\textwidth]{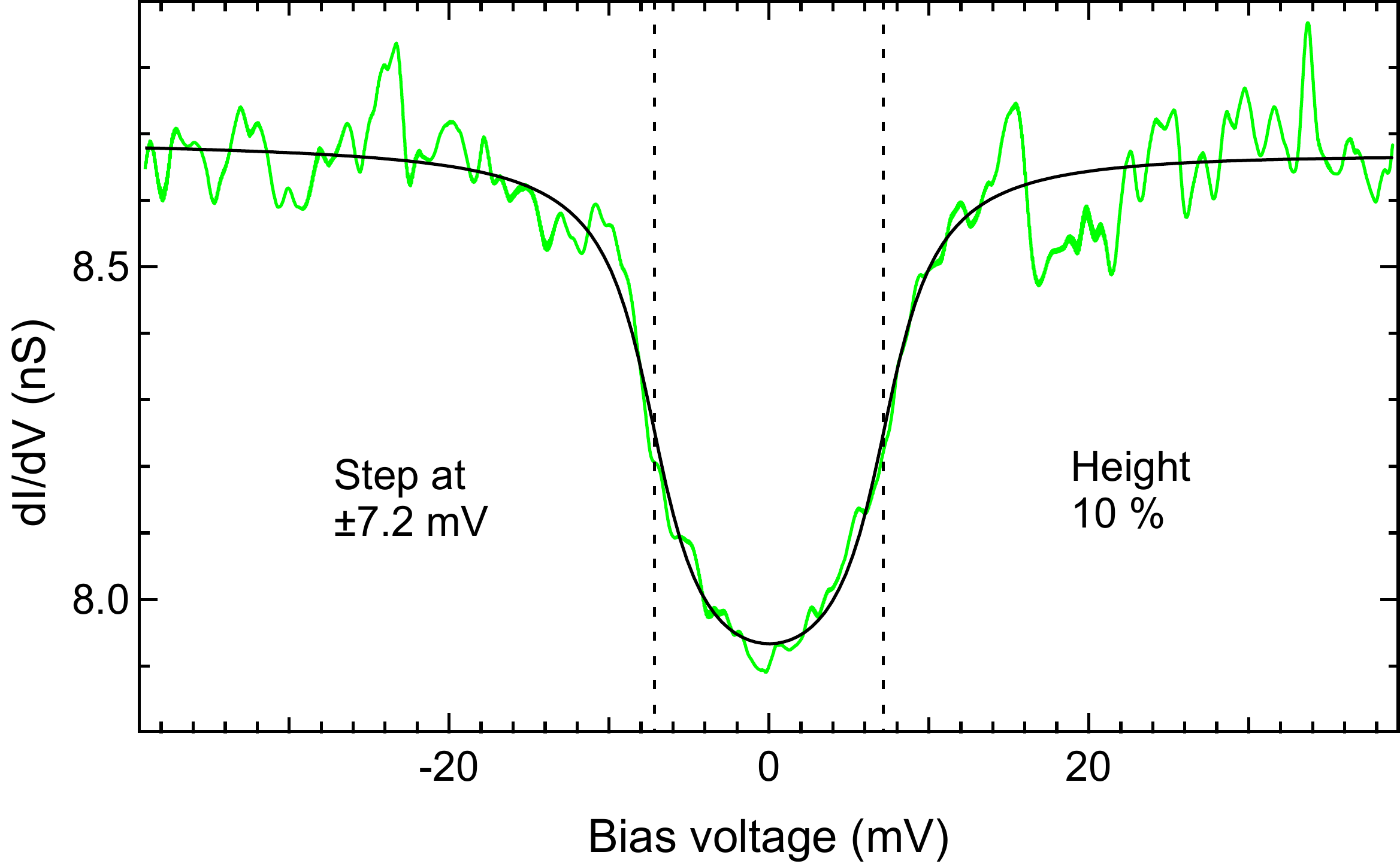}
  \caption{STS spectrum showing inelastic excitations on Gd(tta)$_3$, analogously to Fig.\,\ref{fig7}. The spectrum was recorded in constant-height mode, with the feedback opened at 68\,pA and 70\,mV ($V_\mathrm{mod}=1\,$mV). Black lines are fits using an arctan-step function to determine the step positions at $\pm$7.2(4)\,mV and heights (10$\%$ of the d$I$/d$V$ amplitude).}
 \label{fig7_2}
\end{figure}

Due to the absence of a sizeable magnetic anisotropy in the XMCD data, we expect no inelastic excitations of magnetic origin in the differential conductance spectra recorded on the Gd complexes. However, inelastic tunneling spectra on the Gd(tta)$_3$ complex show a very similar inelastic step as for the Dy-complex (Fig.\,\ref{fig7_2}). Hence, we conclude that these excitations originate from molecular vibrations. The prominent exposure and its decoupling from the substrate render the upper ligand very sensitive to vibrational excitations by electrons.
The absence of inelastic spin excitations could be due to several reasons. First, tunneling into particular $f$ states could be suppressed by symmetry \cite{Bryant15}. However, in our case, the $4f$ states remain largely unperturbed and close to spherical symmetry. The tunneling coupling should therefore not underlie symmetry selection rules due to their shape. Second, the $4f$ electrons are hardly accessible with the tunneling electrons. The tunneling path is probably dominated by the molecular states of the ligand, which are coupled to the $5d$ and $6s$ electrons of the rare-earth ion by coordination bonding. 
An exchange coupling to the $4f$ electrons is therefore likely to be small, such that the magnetic  information can not be accessed directly by electronic transport measurements \cite{Steinbrecher16}.

\section{Conclusions}

Due to their high magnetic moment and anisotropy, Dy(tta)$_3$ molecules are ideal candidates for magnetic data storage when adsorbed on a surface.
The adsorption configuration on Au(111) forces two of the three tta ligands into the surface plane. 
The ligand field imposes a distinguished orientation of the symmetry axis of the $f$ orbitals and direction of magnetization parallel to the surface. Such an easy-axis anisotropy is an ideal situation for creating large anisotropy barriers. The ground state of the Dy$^{3+}$ ion has the maximum projected magnetic moment of $M_J=\pm15/2$, in contrast to Dy-bis(phthalocyaninato) complexes, where an $M_J=\pm13/2$ ground state has been observed \cite{ishikawa03}.

Gd(tta)$_3$ molecules adsorb in exactly the same configuration. However, due to the half-filled $4f$ shell, their anisotropy is vanishingly small, as expected. Features seen in inelastic tunneling spectra appear identically in both systems and are thus attributed to vibrational excitations. No spin excitations are distinguished in the tunneling spectra, which we attribute to the rather shielded nature of the $4f$ electrons.

\section{Acknowledgments}
\begin{acknowledgments}
We gratefully acknowledge discussions with B. W. Heinrich, help in the X-ray absorption experiments by A. Kr\"uger, and funding by the Deutsche Forschungsgemeinschaft through Sonderforschungsbereich 658 and the Emmy-Noether program (CZ 183/1-1) as well as by the European Research Council through the Consolidator Grant ``NanoSpin''. D.R. acknowledges a scholarship by the International Max-Planck Research School ``Functional Interfaces in Physics and Chemistry''. The help of F.\,M.\,F. de Groot with the input files for Cowan's code is gratefully acknowledged.
\end{acknowledgments}

\appendix

\section{Multiplet Calculation of Rare Earth Ions}
\label{calcu}

\begin{figure*}
\center
  \includegraphics[width=0.8\textwidth]{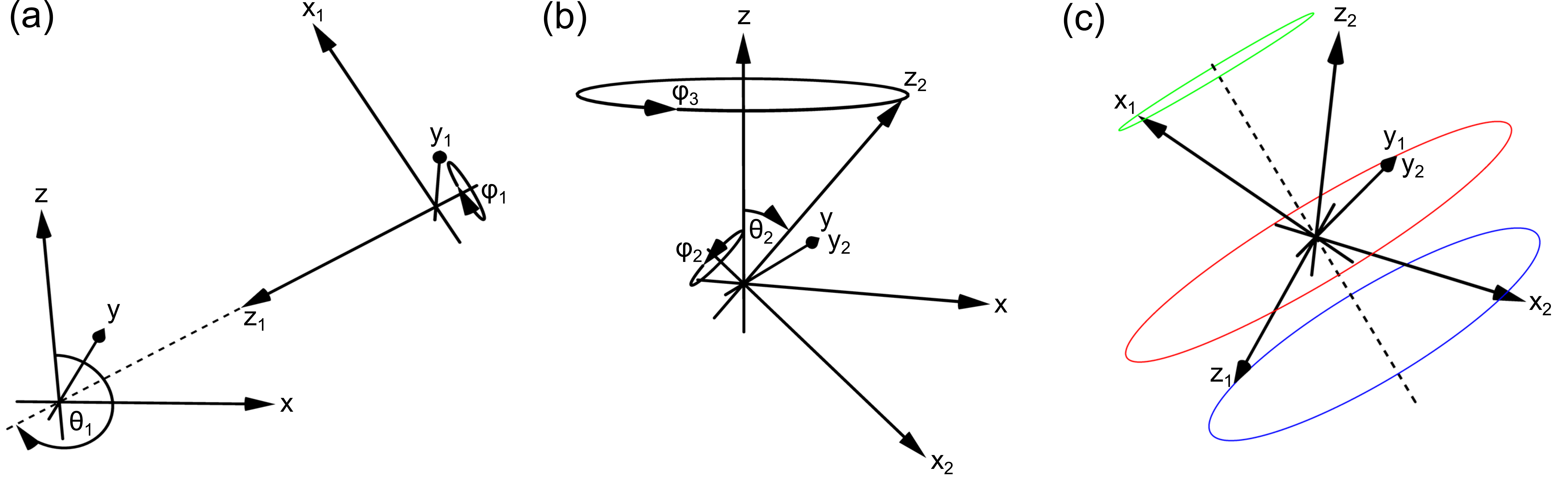}
  \caption{Rotation of the frame of the light (a) and the ion (b) with respect to the frame of the surface, respectively. The $k$ vector of the X rays is parallel to the $z_1$ axis. The symmetry axis of the ion is given by the $z_2$ axis. The orientation of the ions is random with respect to the surface normal given by the $z$ axis. (c) Rotation of the frame of the light with respect to the frame of the ion. Due to the random azimuthal orientation of the ions, the $x_1$, $y_1$, and $z_1$ axes execute cones around the surface normal (dashed line), indicated by green, red, and blue lines, respectively. The opening angle of the cone of the $k$ vector is $2\theta_1$.\label{rotate}}
\end{figure*}

The magnetic moment of rare earth elements stems from their $4f$ electrons. These electrons are localized at the core of the ion and are shielded from direct interactions with the ligands. In weak magnetic fields, the magnetic anisotropy energy and the Zeeman energy are about one order of magnitude smaller than the spin-orbit coupling energy and about two orders of magnitude lower than the Coulomb and exchange energy between the $4f$ electrons. Hence, the Hilbert space corresponding to the atomic Hamilton operator and the one of the Zeeman and anisotropy terms, given in equation (1) of the main manuscript, can be diagonalized separately. The absorption of a photon induces a transition from an initial state $|\alpha J i\rangle=\sum_M a_{iM}|\alpha J M\rangle$ to a final state $|\alpha'\!J'\!f\rangle=\sum_{M'} b_{fM'}|\alpha' \!J'\!M'\rangle$ that can be expressed in the basis of the initial and final state multiplet \cite{GoedkoopPRB88}. Here, $\alpha$ is a placeholder for all quantum numbers that are necessary to fully describe the state.
The operator for an electric dipole transition is given by the position operator and a polarization vector.
To describe situations with varying incidence of the X rays, the frame of the light needs to be rotated with respect to the frame of the surface. Furthermore, the frame of the rare earth ion needs to be rotated, too. 
These transformations are performed by successive rotations around the z, y, and z axes by three Euler angles and are represented by the operators $R_1=R_1(\varphi_1,\theta_1,0)$ and $R_2=R_2(\varphi_2,\theta_2,\varphi_3)$ for rotating the frame of the light and the ion, respectively (see Fig.\,\ref{rotate}).
These rotations are calculated most conveniently using the transformation properties of spherical tensor operators:
\begin{widetext}
\begin{align*}
R_2^\dagger R_1\vec{\varepsilon}\cdot\vec{r}R_1^\dagger R_2
&=r\sqrt{\tfrac{4\pi}{3}}\sum_{q}\varepsilon_{q}R_2^\dagger R_1 Y_{1,q}R_1^\dagger R_2
=r\sqrt{\tfrac{4\pi}{3}}\sum_{qq'}\varepsilon_{q}D_{q'\!q}^{(1)}(\varphi_1,\theta_1,0)R_2^\dagger Y_{1,q'}R_2\\
&=r\sqrt{\tfrac{4\pi}{3}}\sum_{qq'\!q''}\varepsilon_{q}D_{q'\!q}^{(1)}(\varphi_1,\theta_1,0)D_{q'\!q''}^{(1)*}(\varphi_2,\theta_2,\varphi_3)Y_{1,q''}\\
&=rY_{1,1}\sqrt{\tfrac{2\pi}{3}}\Big(\sqrt{1+P_c}\,\mathrm{e}^{-\mathrm{i}(\varphi_1-\varphi_2+\varphi_3)}\left(\mathrm{e}^{\mathrm{i}\varphi_3}\cos\tfrac{\theta_1}{2}\cos\tfrac{\theta_2}{2}+\sin\tfrac{\theta_1}{2}\sin\tfrac{\theta_2}{2}\right)^2\\
&\quad\quad\quad\quad\quad+\sqrt{1-P_c}\,\mathrm{e}^{\mathrm{i}(\varphi_1+\varphi_2-\varphi_3)}\left(\mathrm{e}^{\mathrm{i}\varphi_3}\cos\tfrac{\theta_2}{2}\sin\tfrac{\theta_1}{2}-\cos\tfrac{\theta_1}{2}\sin\tfrac{\theta_2}{2}\right)^2\Big)\\
&+rY_{1,0}\sqrt{\tfrac{\pi}{3}}\Big(\sqrt{1+P_c}\,\mathrm{e}^{-\mathrm{i}\varphi_1}(\cos\theta_2\sin\theta_1-\sin\theta_2(\cos\theta_1\cos\varphi_3+\mathrm{i}\sin\varphi_3))\\
&\quad\quad\quad\quad\quad-\sqrt{1-P_c}\,\mathrm{e}^{\mathrm{i}\varphi_1}(\cos\theta_2\sin\theta_1+\sin\theta_2(-\cos\theta_1\cos\varphi_3+\mathrm{i}\sin\varphi_3))\Big)\\
&+rY_{1,-1}\sqrt{\tfrac{2\pi}{3}}\Big(\sqrt{1+P_c}\,\mathrm{e}^{-\mathrm{i}(\varphi_1+\varphi_2+\varphi_3)}\left(\cos\tfrac{\theta_2}{2}\sin\tfrac{\theta_1}{2}-\mathrm{e}^{\mathrm{i}\varphi_3}\cos\tfrac{\theta_1}{2}\sin\tfrac{\theta_2}{2}\right)^2\\
&\quad\quad\quad\quad\quad+\sqrt{1-P_c}\,\mathrm{e}^{\mathrm{i}(\varphi_1-\varphi_2-\varphi_3)}\left(\cos\tfrac{\theta_1}{2}\cos\tfrac{\theta_2}{2}-\mathrm{e}^{\mathrm{i}\varphi_3}\sin\tfrac{\theta_1}{2}\sin\tfrac{\theta_2}{2}\right)^2\Big)\\
&=r\sum_qc_qY_{1,q}\,,
\end{align*}
%\end{widetext}
with $Y_{1,q}$ the spherical harmonics, $\vec{\varepsilon}=\frac{1}{\sqrt{2}}(\sqrt{1+P_c},0,\sqrt{1-P_c})$ the polarization vector, $|P_c|$ the degree of circular polarization, and $D^{(1)}$ the Wigner D-matrix. $P_c=1$ and $-1$ correspond to right and left circularly polarized light, respectively. Zero degree of circular polarization describes a situation in which the light is fully linearly polarized along the y axis in the frame of the light.

The strength of a particular transition is given by Fermi's golden rule:
%\begin{widetext}
\begin{align*}
S_{\alpha Ji\alpha'\!J'\!f}&=|\langle\alpha Ji|R_2^\dagger R_1\vec{\varepsilon}\cdot\vec{r}R_1^\dagger R_2|\alpha' \!J'\!f\rangle|^2\\
&=\Big|\!\!\sum_{MM'\!q}a_{iM}^*b_{fM'}c_q\langle J'\!M'1q|JM\rangle\langle\alpha J||Y_1||\alpha' \!J'\rangle\langle\alpha J|r|\alpha' \!J'\rangle\Big|^2.
\end{align*}
Applying the Wigner-Eckart theorem we can separate the matrix element into a factor that does not depend on the magnetic quantum numbers and an angular and helicity dependent part containing Clebsch-Gordan coefficients.
Within this approach the orientation-dependent part of the radial matrix element is neglected. It therefore does not include the contributions of the asymmetry of the charge and spin density distributions. This is justified here since for the isotropic case of Gd$^{3+}$ these contributions are vanishing and for Dy$^{3+}$ they are smaller than 10\% (see discussion in Appendix\,\ref{effmom}).

The energy separation of the individual levels of one multiplet is much smaller than the life-time broadening and the energy resolution of the experiment and is in the same energy range as the temperature. Thus, the observed transitions between two multiplets are a sum over the final states and a Boltzmann-weighted sum over the initial states:
\begin{align*}
S_{\alpha J\alpha'\!J'}&=\frac{1}{Z}\sum_i\mathrm{e}^{-\frac{E_i}{kT}}\sum_fS_{\alpha Ji\alpha'\!J'\!f}\\
&=\frac{1}{Z}\sum_i\mathrm{e}^{-\frac{E_i}{kT}}\,\Big|\!\!\sum_{MM'\!q}a_{iM}^*c_q\langle J'M'1q|JM\rangle\Big|^2|\langle\alpha J||Y_1||\alpha'\!J'\rangle|^2|\langle\alpha J|r|\alpha'\!J'\rangle|^2,
\end{align*}
\end{widetext}
with $Z$ the partition function. Since the sum of the final states is carried out over the complete orthogonal subspace of the multiplet, it is independent of the expansion coefficients.
\begin{table*}
\begin{tabular}{ccccccccccc}
&$J$&$D\,$(meV)&$\kappa_1$&$\kappa_2$&$\kappa_3$&$\alpha_{3d}$&$\alpha_{4f}$&$2\Gamma_{5/2}\,$(eV)&$2\Gamma_{3/2}\,$(eV)&$q_{3/2}$\\
Dy(tta)$_3$&$15/2$&-0.68&0.77&1.00&0.80&0.98&1.01&0.56&0.77&8\\
Gd(tta)$_3$&$7/2$&-0.02&0.88&0.98&0.84&0.98&0.97&0.49&0.81&14\\
\end{tabular}
  \caption{Parameters used to fit all experimental spectra simultaneously, {\it i.e.}, scaling parameters for the reduction with respect to the values calculated within the Hartree-Fock approach and parameters applied to broaden the transition lines.}
  \label{multi}
\end{table*}
The expansion coefficients of the initial state and the eigenvalues are obtained by diagonalizing the Hamiltonian given in equation (1) of the main manuscript. The angular- and helicity-independent parts of the matrix element and the corresponding transition energies are calculated with Cowan's code \cite{Cowan81} in the version provided with the CTM4XAS program by de Groot ~\cite{Stavitski10}. Three multiplets $J'=J-1,J,J+1$ are accessible from the ground state multiplet corresponding to three spectra. These spectra are obtained by an atomic calculation and then weighted according to the equation above. Averaging over $\varphi_3$ was carried out to account for the random orientation of the ions with respect to the surface normal.

Within the atomic calculations, radial wave functions are obtained by a Hartree-Fock approach. To account for the underestimation of electron correlation, the Hartree-Fock-calculated values are scaled down by reduction factors such that the spectra fit to the experimental ones.
Three scaling factors $\kappa_1,\kappa_2,$ and $\kappa_3$ are used for the $4f4f$ Slater integrals, the $3d4f$ direct Slater integrals, and the $3d4f$ exchange Slater integrals, respectively. Two scaling factors $\alpha_{3d}$ and $\alpha_{4f}$ are used for the spin-orbit coupling of the $3d$ and $4f$ electrons, respectively. The same factors were used for the initial and final state. Transition energies at the $M_5$ edge are broadened with a Lorentz profile of width $2\Gamma_{5/2}$. At the $M_4$ edge, a Fano profile of width $2\Gamma_{3/2}$ and asymmetry parameter $q_{3/2}$ is used to account for mixing with $3d_{5/2}$ transitions to the continuum~\cite{Thole85}. The resulting spectra are then convoluted with a Gauss profile accounting for the experimental resolution of $\sigma=160$\,meV. Simulations have been performed by first calculating the Dy $M_{4,5}$ XA spectra of a Dy$^{3+}$ ion without ligand field separately for each of the $\Delta J = 0,\pm1$ dipole-transitions. The mean squared deviation between the simulated and experimental spectra was minimized for all experimental spectra simultaneously using the anisotropy parameter determined from the field-dependent integrated XMCD.
The resulting parameters are given in Tab.\,\ref{multi}. The simultaneous fit is only possible when the symmetry axis of the $f$ orbitals lies parallel to the surface and the ground state is $M_J=\pm15/2$, as discussed in Sec.\,\ref{elstruc}.

\section{Calculated Spectra of Individual Kramers Doublets}
\label{kramer}

\begin{figure}
\begin{tabular}{c}
  \includegraphics[width=0.48\textwidth]{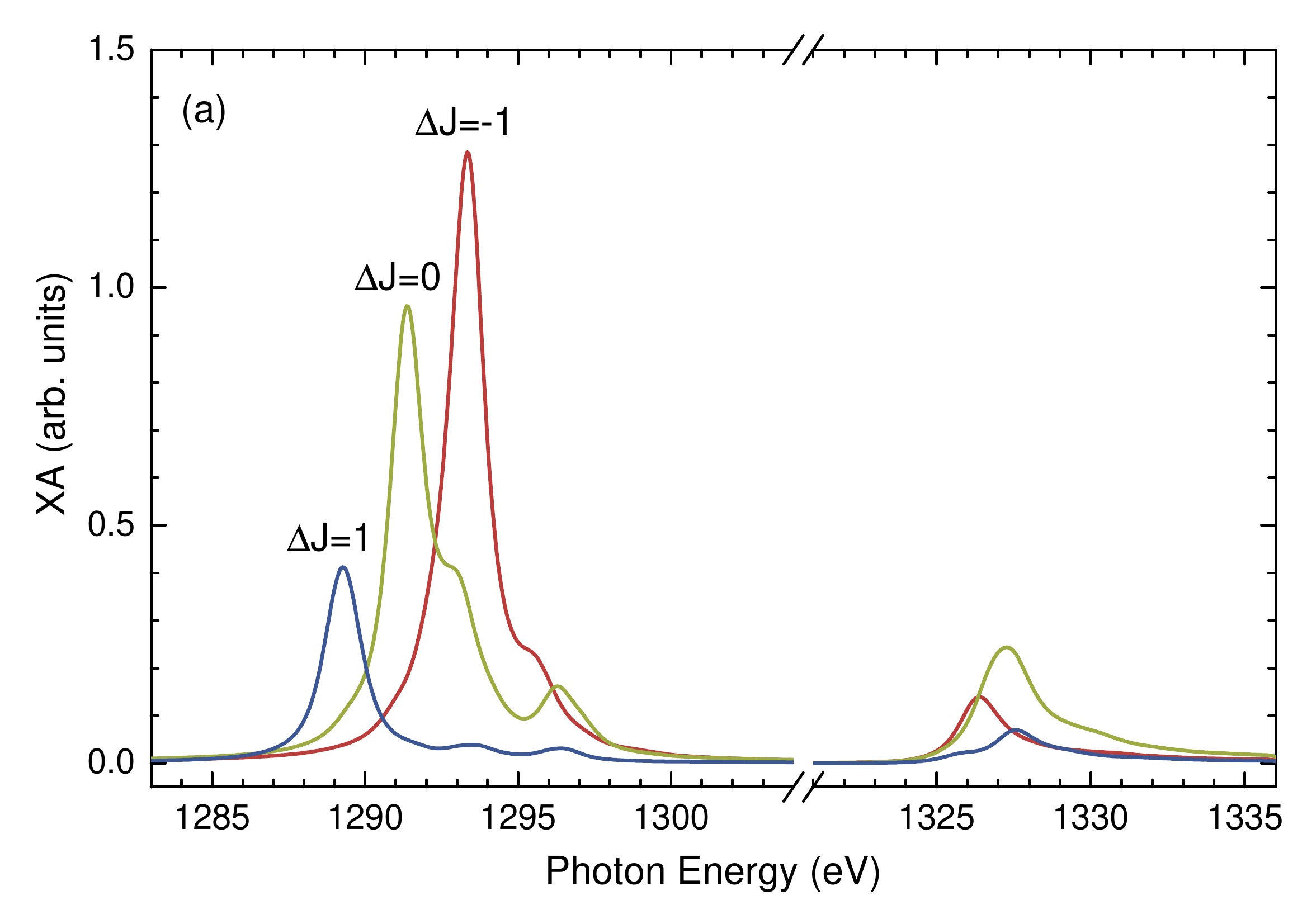}\\
  \includegraphics[width=0.45\textwidth]{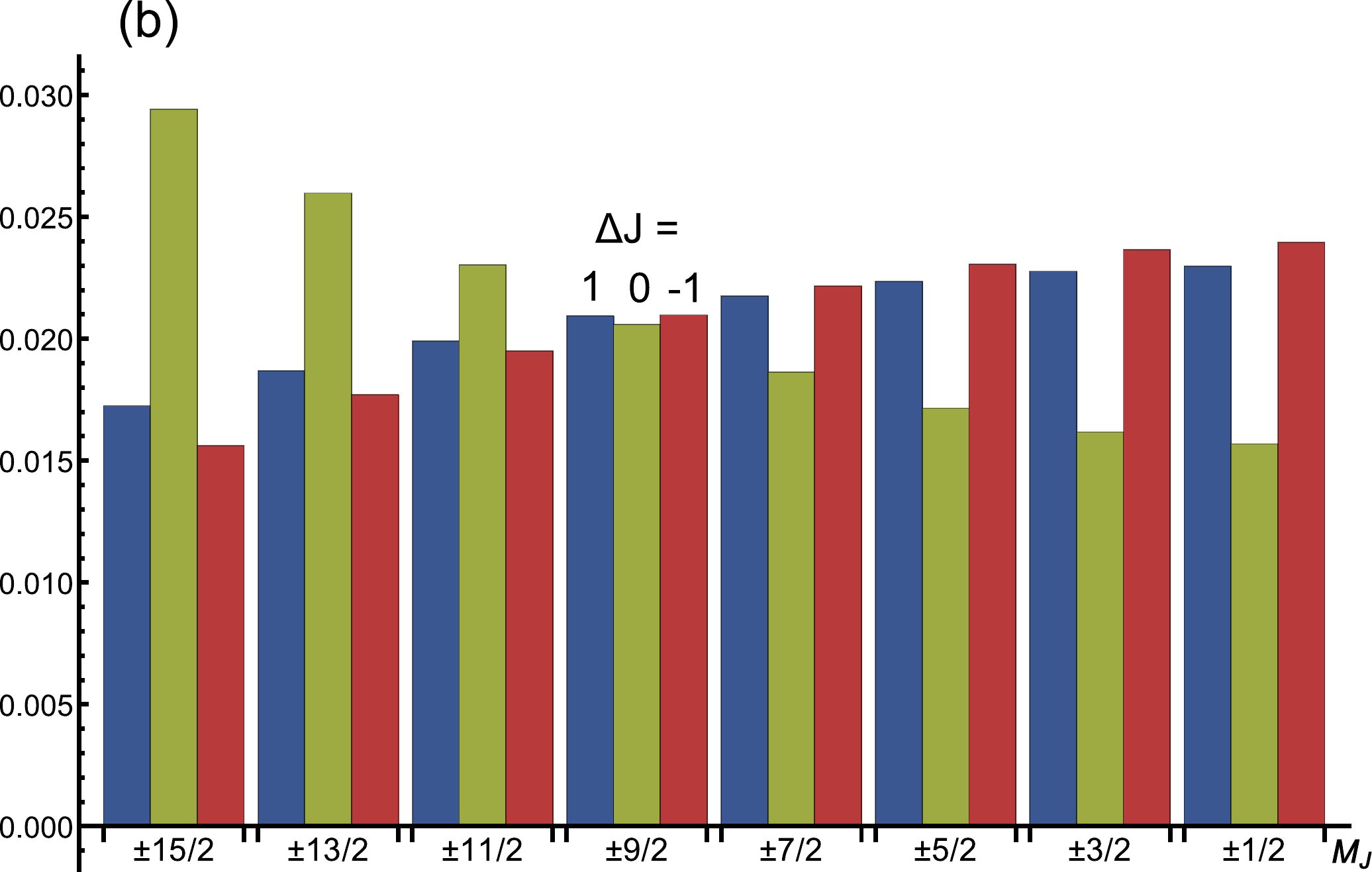}\\
\end{tabular}
  \caption{(a) Calculated Dy $M_{4,5}$ XA spectra in the absence of a ligand field for the three dipole-allowed transitions using the parameters shown in Tab.\,\ref{multi}. (b) Helicity- and orientation-dependent weighting factors of the transitions into the three final state multiplets for the individual Kramers doublets and a situation in which the symmetry axis of the $f$ orbitals is parallel and the incidence of the X rays perpendicular to the surface.}
  \label{contribu}
\end{figure}

In Figure\,\ref{contribu}(a), calculated Dy $M_{4,5}$ XA spectra for the three dipole-allowed transitions into the final state multiplets $J'=$ 13/2, 15/2, and 17/2 are shown. These spectra are calculated with Cowan's code \cite{Cowan81} in the absence of a ligand field using the parameters shown in Tab.\,\ref{multi}. The final spectrum is calculated as the sum of these three contributions using weighting factors that depend on the initial state, the helicity, and the orientation of the light. In Figure\,\ref{contribu}(b) these weighting factors are shown for the individual Kramers doublets and a situation in which the symmetry axis of the $f$ orbitals is parallel and the incidence of the X rays perpendicular to the surface. The ratio between contributions with $\Delta J=-1$ and $+1$ shows a slight variation for the different doublets. The contribution of the transition with $\Delta J=0$ with respect to the ones with $\Delta J=\pm 1$ shows a significant variation.

\section{Sum-Rule Analysis of XMCD Spectra}
\label{effmom}

\begin{table}
\begin{tabular}{|cc|@{~~~}c@{~~~}cc|}
 \hline
 & & $\langle S_\alpha^\mathrm{eff}\rangle$ & $\langle L_\alpha\rangle$ & $2\langle S_\alpha^\mathrm{eff}\rangle+\langle L_\alpha\rangle$ \\\hline
 Dy(tta)$_3$ & $25^\circ$ & 1.03(08) & 2.74(18) & 4.79(24) \\
    & $35^\circ$ & 0.80(06) & 2.24(15) & 3.84(19) \\
    & $90^\circ$ & 0.08(04) & 0.29(06) & 0.45(10) \\\hline
 Gd(tta)$_3$ & $20^\circ$ & 3.03(18) & -0.11(16) & 5.95(39) \\
    & $90^\circ$ & 2.99(23) & -0.15(24) & 5.82(52) \\
 \hline
\end{tabular}
  \caption{Expectation values of the effective spin, the orbital, and the effective total magnetic moment determined from a sum-rule analysis of the XA and XMCD spectra presented in Fig.\,\ref{Dy-xmcd} and Fig.\,\ref{Gd-xas}(c) and (d). In addition, the values from Dy $M_{4,5}$ spectra recorded at $35^\circ$ incidence under the same conditions are shown.}
  \label{sumrule}
\end{table}

In Table\,\ref{sumrule} the values of the effective spin, the orbital, and the total magnetic moment determined from a sum-rule analysis \cite{Carra93} of the Dy and Gd $M_{4,5}$ XA and XMCD spectra presented in Fig.\,\ref{Dy-xmcd} and Fig.\,\ref{Gd-xas}(c) and (d) are shown, where $n_h=5$ and 7 was used as the number of $f$ holes for Dy$^{3+}$ and Gd$^{3+}$, respectively.

For Gd(tta)$_3$ the orbital moment is almost zero as expected for a $4f^7$ system. Since the spin density distribution is isotropic, in such a case the intra-atomic magnetic dipole operator $T_\alpha$ ($\alpha=x,y,z$) is vanishing and $\langle S_\alpha^\mathrm{eff}\rangle=\langle S_\alpha\rangle$ \cite{Carra93,teramura1996}. Extrapolating the experimentally determined value of the expectation value of the total magnetic moment to full saturation by using the value obtained by fitting the model (Fig.\,\ref{mfitGd}) yields $6.3(5)\,\mu_B$ for both incidence angles. This value is slightly lower than the saturated magnetic moment of a free Gd$^{3+}$ ion of $7\,\mu_B$. The deviation can be attributed to shortcomings of the sum rules due to mixing of the $3d_{3/2}$ and $3d_{5/2}$ initial states which leads to a reduction of 5\% for Gd$^{3+}$ \cite{teramura1996}.

\begin{figure}
\center
  \includegraphics[width=0.48\textwidth]{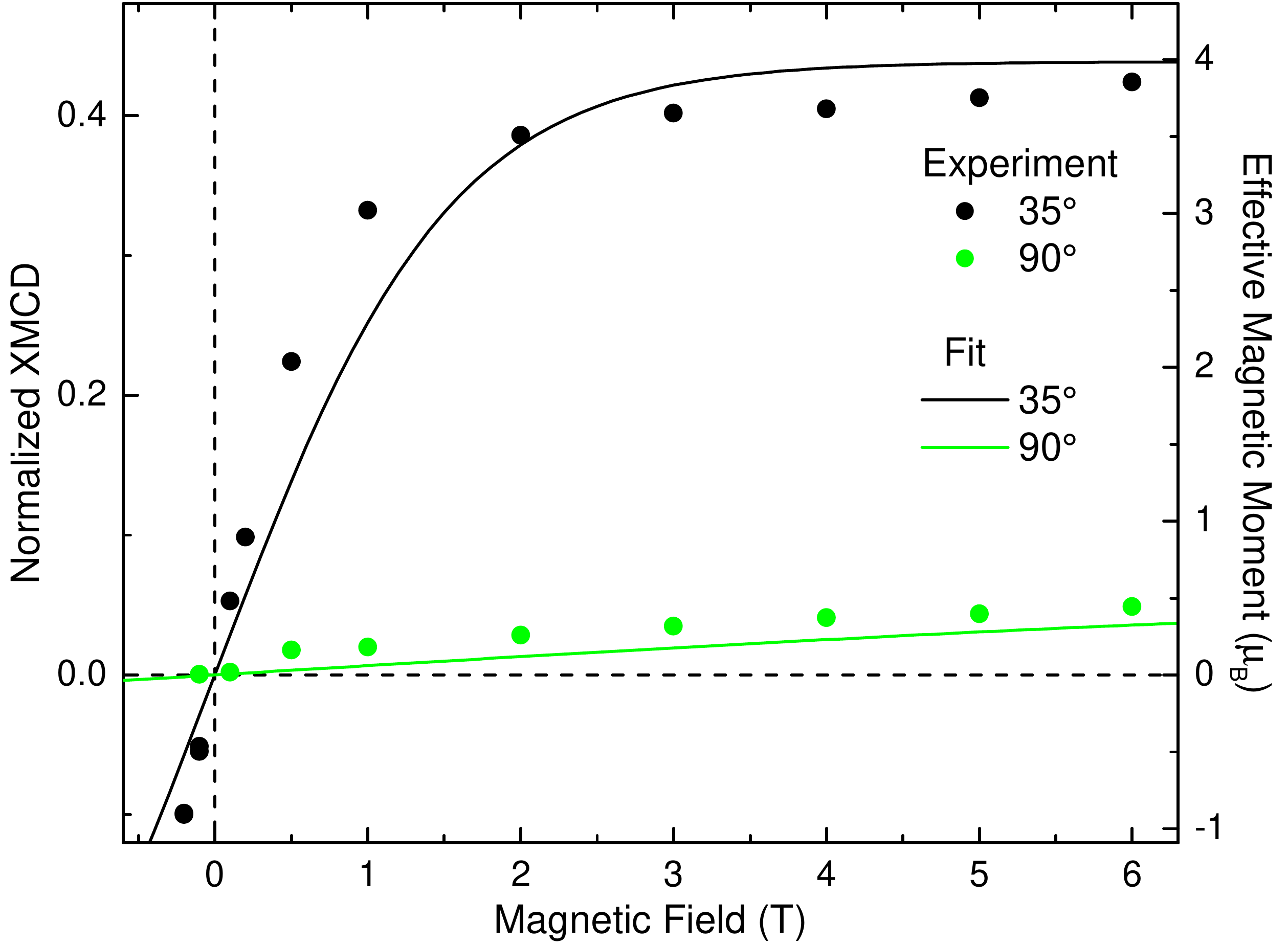}
  \caption{Magnetic field dependence of the integrated Dy-$M_5$ XMCD signal at $35^\circ$ and $90^\circ$ incidence angles. The experimental data is the same as the one shown in Fig.\,\ref{mfit}(a). The right y axis displays the experimentally determined effective magnetic moment derived from a sum-rule analysis (see Appendix\,\ref{effmom}).
  }\label{fitDy-posD-magB}
\end{figure}

Extrapolating the expectation value of the total magnetic moment of Dy(tta)$_3$ at $35^\circ$ incidence to full saturation by using the value obtained by fitting the model (Fig.\,\ref{fig4}) yields $7.1(4)\,\mu_B$. This value is lower than the saturated magnetic moment of a free Dy$^{3+}$ ion of $10\,\mu_B$. The deviation maybe be attributed to shortcomings of the sum rules. Mixing of the $3d_{3/2}$ and $3d_{5/2}$ initial states leads to a reduction of 8\% for Dy$^{3+}$ \cite{teramura1996}. In addition, there is a reduction due to the asymmetry of the spin-density distribution according to $\langle S_z^\mathrm{eff}\rangle=\langle S_z\rangle+3\langle T_z\rangle$ \cite{Carra93} with $\langle T_z\rangle/\langle S_z\rangle=-0.053$ \cite{teramura1996} and $\langle T_z\rangle/\langle S_z\rangle=-2\langle T_x\rangle/\langle S_x\rangle=-2\langle T_y\rangle/\langle S_y\rangle$.
This means that the observed expectation value of the total magnetic moment along the symmetry axis of the $f$ orbitals is reduced by about 8\% and perpendicular to it is enhanced by about 3\%.
The remaining discrepancies may be ascribed to limitations of the simplified model in the description of the ligand field.
Alternatively, the reduction could be explained by an arrangement of the energy levels within the ground state multiplet in which the $M_J=\pm15/2$ doublet is not the one lowest in energy. However, such a situation can be excluded here due to the steep increase of the XMCD signal at $35^\circ$ (Fig.\,\ref{fig4}) and the line shape of the Dy XA spectrum shown in Fig.\,\ref{mj}.

\section{Model with Easy-Plane Anisotropy}
\label{posD}

The magnetic behavior of Dy(tta)$_3$ on Au(111) cannot be modeled satisfactorily assuming an easy-plane magnetic anisotropy and an orientation of the symmetry axis of the $f$ orbitals perpendicular to the surface. Figure\,\ref{fitDy-posD-magB} shows the best fit of the magnetic field-dependent XMCD data of Dy(tta)$_3$ to such a model. The fit yields a diverging positive value for $D$. Compared to Fig.\,\ref{mfit}(a), clear systematic deviations are observed for both field angles, which are also reflected by a 3.6 times higher mean squared deviation. An easy-plane anisotropy for Dy(tta)$_3$ on Au(111) can therefore be excluded.

\section{XMCD Temperature Calibration}

\begin{figure}
\center
  \includegraphics[width=0.48\textwidth]{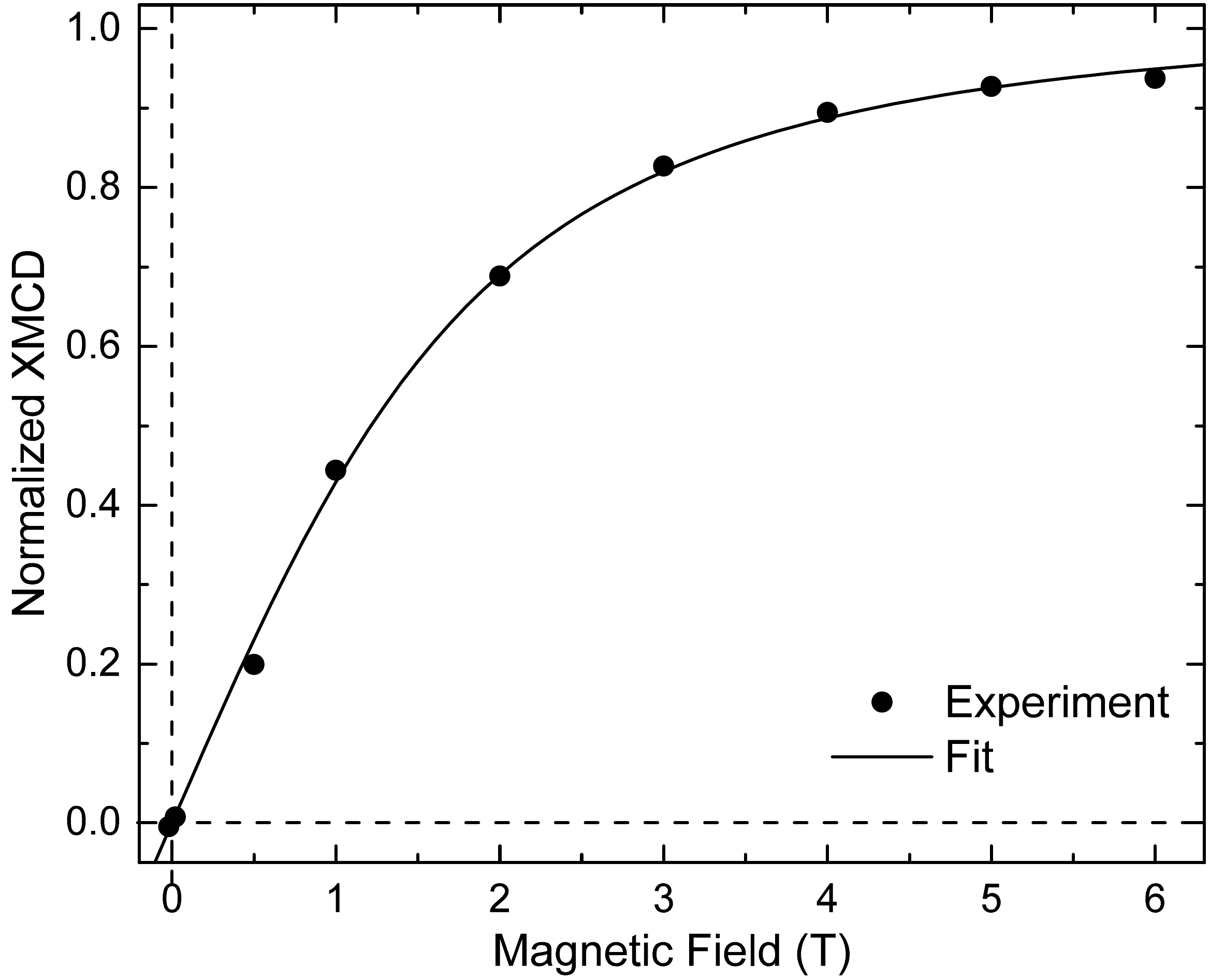}
  \caption{Integrated Gd $M_5$ XMCD signal of a Gd(III) sulfate bulk-reference sample measured as a function of external magnetic field. The solid black line is a Brillouin function fit to the experimental data (see text). The XMCD signal has been normalized to its saturated value.
  }\label{saltGd}
\end{figure}

For the XMCD experiments, a single-crystalline Au(111) substrate was mounted by an $0.3\,$mm Ta foil spot-welded to a Mo Omicron sample plate. The sample plate was screwed to a copper block (shuttle) on the sample holder and coupled to a liquid-He cryostat, which was pumped to about 25~mbar. The temperature was measured at the sample holder by means of a calibrated diode. The sample was shielded by the surrounding bath cryostat that was cooling the superconducting coils. To get an estimate of the difference between the temperature measured by the diode and the temperature at the sample surface, a reference measurement was carried out using Gd(III) sulfate octahydrate (Gd$_2$(SO$_4$)$_3\cdot$8H$_2$O) powder ground into indium foil mounted on a sample plate. In Figure\,\ref{saltGd}, the integrated Gd $M_5$ XMCD signal is plotted as a function of external magnetic field at a temperature of $4.4(1)$\,K as measured with the diode. The experimental data is fitted with a Brillouin function with $J=7/2$ and $g=2$. The temperature determined by the fit of $T=4.3$~K is the same within error as measured at the sample holder by the diode. We thus conclude that the temperature at the sample surface can be well approximated with the one at the diode.

\section{Molecular Coverage}

The determination of the molecular coverage in the XA experiments was carried out by means of the O $K$ edge jump. As reference, half a layer of atomic oxygen on a reconstructed Cu(001) single crystal was used \cite{Sorg06}, displaying an O $K$ edge jump of 5\% and an atom density of $7.7\,$ atoms/nm$^2$. Such a sample can be prepared reliably using the self-terminated oxidation of Cu(001) at $T=500\,$K, a pressure of $2\times10^{-6}\,$mbar oxygen with a dose of about 1200 Langmuir. A ratio of 4.5 between the Au(111) and Cu(001) XA background signal at 500\,eV was determined by XA measurements of the two substrates in identical experimental geometry. The packing density of the Dy(tta)$_3$ complexes of $1.9(1)$ molecules/nm$^2$ is determined from $100\times100$\,nm STM topography images of single terraces like the ones shown in  Fig.\,\ref{fig1}. The experimentally observed O $K$ edge jump of 0.1\% results in a coverage of about 0.2\,ML.

\end{document}